\documentclass[sigconf, nonacm]{acmart}

\AtBeginDocument{%
	\providecommand\BibTeX{{%
			\normalfont B\kern-0.5em{\scshape i\kern-0.25em b}\kern-0.8em\TeX}}}

\newcommand\vldbdoi{XX.XX/XXX.XX}
\newcommand\vldbpages{XXX-XXX}
\newcommand\vldbvolume{15}
\newcommand\vldbissue{12}
\newcommand\vldbyear{2022}

\newcommand\vldbtitle{\shorttitle}
\newcommand\vldbavailabilityurl{https://github.com/milvus-io/milvus/tree/2.0}
\newcommand\vldbpagestyle{empty}

\usepackage{multirow}
\usepackage{subcaption}
\usepackage{url}
\usepackage{docmute}
\usepackage{fancyhdr,graphicx,amsmath}

\usepackage{amssymb}
\usepackage{algorithm, algorithmicx}
\usepackage{algpseudocode}
\usepackage{booktabs}
\usepackage{epstopdf}
\usepackage{flushend}
\usepackage[utf8]{inputenc}
\usepackage{graphicx}
\usepackage{enumerate}
\usepackage{listings}
\usepackage{fancyvrb}
\usepackage{balance}

\newcommand{\name}{$\mathsf{Manu}$}

\newcommand{\stitle}[1]{\vspace*{0.4em}\noindent{\bf #1:\/}}
\newcommand{\sntitle}[1]{\vspace*{0.4em}\noindent{\bf  #1\/}}
\newcommand{\sstitle}[1]{\vspace*{0.4em}\noindent{\bf #1\/.}}

\newcommand{\squishlist}{
	\begin{list}{$\bullet$}
		{ \setlength{\itemsep}{1pt}
			\setlength{\parsep}{1pt}
			\setlength{\topsep}{2.5pt}
			\setlength{\partopsep}{0.5pt}
			\setlength{\leftmargin}{1em}
			\setlength{\labelwidth}{1em}
			\setlength{\labelsep}{0.6em}
		}
	}
	\newcommand{\squishend}{
	\end{list}
}

\begin{document}
\fancyhead{}

\title{Manu: A Cloud Native Vector Database Management System}

\author{Rentong Guo$^\dagger$$^*$, Xiaofan Luan$^\dagger$$^*$, Long Xiang$^\ddagger$$^*$, Xiao Yan$^\ddagger$$^*$, Xiaomeng Yi$^\dagger$$^*$, Jigao Luo$^{\dagger \S}$ \and Qianya Cheng$^\dagger$, Weizhi Xu$^\dagger$, Jiarui Luo$^\ddagger$, Frank Liu$^\dagger$, Zhenshan Cao$^\dagger$, Yanliang Qiao$^\dagger$, Ting Wang$^\dagger$ \and Bo Tang$^\ddagger$  Charles Xie$^\dagger$}
\affiliation{
	\institution{$^\dagger$Zilliz\\
		$^\ddagger$Department of Computer Science and Engineering, Southern University of Science and Technology \\
		$^\S$Technical University of Munich\\
		$^\dagger$\{firstname.lastname\}@zilliz.com \\
		$^\ddagger$\{xiangl3@mail., yanx@, 11911419@mail., tangb3@\}sustech.edu.cn, 
		$^\S$jigao.luo@tum.de}
	\city{}
	\country{}
}

\begin{abstract}


With the development of learning-based embedding models, embedding vectors are widely used for analyzing and searching unstructured data. As vector collections exceed billion-scale, fully managed and horizontally scalable vector databases are necessary. In the past three years, through interaction with our 1200+ industry users, we have sketched a vision for the features that next-generation vector databases should have, which include long-term evolvability, tunable consistency, good elasticity, and high performance.

We present \name, a cloud native vector database that implements these features.
It is difficult to integrate all these  features if we follow traditional DBMS design rules. As most vector data applications do not require complex data models and strong data consistency, our design philosophy is to relax the data model and consistency constraints in exchange for the aforementioned features. Specifically, \name~firstly exposes the write-ahead log (WAL) and binlog as backbone services. Secondly, write components are designed as log publishers while all read-only analytic and search components are designed as independent subscribers to the log services. Finally, we utilize multi-version concurrency control (MVCC) and a delta consistency model to simplify the communication and cooperation among the system components. These designs achieve a low coupling among the system components, which is essential for elasticity and evolution. We also extensively optimize \name~for performance and usability with hardware-aware implementations and support for complex search semantics.
\name~has been used for many applications, including, but not limited to, recommendation, multimedia, language, medicine and security. We evaluated \name~in three typical application scenarios to demonstrate its efficiency, elasticity, and scalability.

\end{abstract}

\maketitle

\renewcommand{\thefootnote}{\fnsymbol{footnote}}
\footnotetext[1]{Co-first-authors are ordered alphabetically. \\
$^\ddagger$ Work done while working with Zilliz, correspondence to Bo Tang.} 
\renewcommand{\thefootnote}{ \arabic{footnote}}


\pagestyle{\vldbpagestyle}
\begingroup\small\noindent\raggedright\textbf{PVLDB Reference Format:}\\
Rentong Guo, Xiaofan Luan, Long Xiang, Xiao Yan, Xiaomeng Yi, Jigao Luo, Qianya Cheng, Weizhi Xu, Jiarui Luo, Frank Liu, Zhenshan Cao, Yanliang Qiao, Ting Wang, Bo Tang, and Charles Xie.
\vldbtitle. PVLDB, \vldbvolume(\vldbissue): \vldbpages, \vldbyear.\\
\href{https://doi.org/\vldbdoi}{doi:\vldbdoi}
\endgroup
\begingroup
\renewcommand\thefootnote{}\footnote{\noindent
This work is licensed under the Creative Commons BY-NC-ND 4.0 International License. Visit \url{https://creativecommons.org/licenses/by-nc-nd/4.0/} to view a copy of this license. For any use beyond those covered by this license, obtain permission by emailing \href{mailto:info@vldb.org}{info@vldb.org}. Copyright is held by the owner/author(s). Publication rights licensed to the VLDB Endowment. \\
\raggedright Proceedings of the VLDB Endowment, Vol. \vldbvolume, No. \vldbissue\ %
ISSN 2150-8097. \\
\href{https://doi.org/\vldbdoi}{doi:\vldbdoi} \\
}\addtocounter{footnote}{-1}\endgroup

\ifdefempty{\vldbavailabilityurl}{}{
\vspace{.3cm}
\begingroup\small\noindent\raggedright\textbf{PVLDB Artifact Availability:}\\
The source code, data, and/or other artifacts have been made available at \url{https://github.com/milvus-io/milvus/tree/2.0}.
\endgroup
}

\section{Introduction}\label{sec:intro}

According to IDC, unstructured data, such as text, images, and video, took up about 80\% of the 40,000 exabytes of new data generated in 2020, their percentage keeps rising due to the increasing amount of human-generated rich media~\cite{ref:idc}. With the rise of learning-based embedding models, especially deep neural networks, using embedding vectors to manage unstructured data has become commonplace in many applications such as e-commerce, social media, and drug discovery ~\cite{mikolov2013distributed,lecun1995convolutional,scarselli2008graph}. A core feature of these applications is that they encode the semantics of unstructured data into a high-dimensional vector space.
Given the representation power of embedding vectors, operations like recommendation, search, and analysis can be implemented via similarity-based vector search.
To support these applications, many specialized vector databases are built to manage vector data~\cite{wang2021milvus,weaviate,vespa,vald,qdrant,pinecone}.

In 2019, we open sourced Milvus~\cite{wang2021milvus}, our previous vector database, under the LF AI \& Data Foundation. Since then, we collected feed-backs from more than 1200 industry users and found that some of the design principles adopted by Milvus are not suitable. Milvus followed the design principles of relational databases, which are optimized for either transaction~\cite{li2018design} or analytical~\cite{wang2021milvus} workloads, and focused on functionality supports (e.g., attribute filtering and multi-vector search) and execution efficiency (e.g., SIMD and cache optimizations). However, vector database applications have different requirements in the following three aspects, which motivates us to restructure \name{} from scratch with focuses on a cloud-native architecture.


\squishlist

\item  \textbf{Support for complex transactions is not necessary.} Instead of decomposing entity representations into different fields or tables, learning-based models encode complex and hybrid data semantics into a single vector. As a result, multi-row or multi-table transactions are not necessary; row-level ACID is sufficient for the majority of vector database applications.



\item  \textbf{A tunable performance-consistency trade-off is important.} Different users have different consistency requirements; some users prefer high throughput and eventual consistency, while others require some level of guaranteed consistency, i.e., newly inserted data should be visible to queries either immediately or within a pre-configured time. Traditional relational databases generally support either strong consistency or eventual consistency; there is little to no room for customization between these two extremes. As such, \textit{tunable} consistency is a crucial attribute for cloud-native vector databases.


\item  \textbf{High hardware cost calls for fine-grained elasticity.} Some vector database operations (e.g., vector search and index building) are computationally intensive, and hardware accelerators (e.g. GPUs or FPGAs) and/or a large working memory are required for good performance. However, depending on application types, workload differs amongst database functionalities. Thus, resources can be wasted or improperly allocated if the vector database does not have fine-grained elasticity. This necessitates a careful decoupling of functional and hardware layers; system-level decoupling such as separation of read and write logic is insufficient, elasticity and resource isolation should be managed at the functionalities level rather than the system level.

\squishend


In summary, modern vector databases should have tunable consistency, functionality-level decoupling, and per-component scalability. Following the design principles of traditional relational databases makes achieving these design goals extremely difficult, if not impossible. A key opportunity for achieving these design goals lies in the potential for relaxing transaction complexity.


\name~follows the ``log as data'' paradigm. Specifically, \name~structures the entire system as a group of log publish/subscribe micro-services. The write-ahead log (WAL) and inter-component messages are published as ``logs", i.e., durable data streams that can be subscribed. Read-side components, such as search and analytical engines, are all built as log subscribers. This architecture provides a simple yet effective way to decouple system functionalities; it enables the decoupling of read from write, stateless from stateful, and storage from computing. Each log entry is assigned a global unique timestamp, and special log entries called time-tick (similar to watermarks in Apache Flink~\cite{flink}) are periodically inserted into each log channel signaling the progress of event-time for log subscribers. The timestamp and time-tick form the basis of the tunable consistency mechanism and multi-version consistency control (MVCC). To control the consistency level, a user can specify a tolerable time lag between a query's timestamp and the latest time-tick consumed by a subscriber.


Additionally, we extensively optimize \name~for performance and usability. \name~supports various indexes for vector search, including vector quantization~\cite{ge2013optimized,akbarinia2007best,guo2016quantization,wu2017multiscale}, inverted index~\cite{babenko2014inverted}, and proximity graphs~\cite{fu2019fast}. In particular, we tailor the implementations to better utilize the parallelization capabilities of modern CPUs and GPUs along with the improved read/write speeds of SSDs over HDDs. \name~ also integrates refactored functionalities from Milvus~\cite{wang2021milvus}, such as attribute filtering and multi-vector search. Moreover, build a visualization tool that allows users to track the performance of \name~in real time and include an auto-configuration tool that recommends indexing algorithm parameters using machine learning. 

To summarize, this paper makes the following contributions:

\squishlist

\item We summarize lessons learned from communicating with over 1200 industry users over three years. We shed light on typical application requirements of vector databases and show how they differ from those of traditional relational databases. We then outline the key design goals that vector databases should meet.

\item We introduce \name's key architectural designs as a cloud native vector database, building around the core design philosophy of relaxing transaction complexity in exchange for tunable consistency and fine-grained elasticity.

\item We present important usability and performance-related enhancements, e.g., high-level API, a GUI tool, automatic parameter configuration, and SSD support.


\squishend


The rest of the paper is organized as follows. Section~\ref{sec:overall} provides background on the requirements and design goals of vector databases. Section~\ref{sec:arch} dives deep into \name's design. Section~\ref{sec:feature} highlights the key features for usability and performance. Section~\ref{sec:case} discusses representative use cases for \name. Section~\ref{sec:related} review related works. Section~\ref{sec:conclusion} concludes the paper and outlines future work.

\section{background and motivation} \label{sec:overall}

Consider video recommendation as a typical use case of vector databases. The goal is to help users discover new videos based on their personal preferences and previous browsing history. Using machine learning models (especially deep neural networks), features of users and videos, such as search history, watch history, age, gender, video language, and tags are converted to embedding vectors. These models are carefully designed and trained to encode the similarity between user and video vectors into a common vector space. Recommendation is conducted by retrieving candidate videos from the collection of video vectors via similarity scores with respect to the specified user vector. The system also needs to handle updates to vectors when new videos are updated, some videos are deleted and the embedding model is changed.

Video recommendation and other applications of vector databases can involve hundreds of billions of vectors with daily growth at hundred-million scale, and serve million-scale queries per second (QPS). Existing DBMSs (e.g., relational databases~\cite{ref:mysql,ref:postgresql}, NoSQL~\cite{zaharia2010spark,thusoo2009hive}, NewSQL~\cite{huang2020tidb,taft2020cockroachdb}) were not built to manage vector data on that scale. Moreover, the underlying data management requirements of their applications differ greatly from vector database applications.


First, when compared with relational databases, both the architecture and theory of vector databases are far from mature. A key reason for this is that AI- and data-driven applications are still in a state of constant evolution, thereby necessitating continued architectural and functionality changes to vector databases as well.



Second, complex transactions are unnecessary for vector databases. In the above example, the recommendation system encodes all semantic features of users and videos into standalone vectors as opposed to multi-row or multi-column entity fields in a relational database. As a result, row-level ACID is sufficient; multi-table operations (such as joins) are inessential.



Third, vector database applications need a flexible performance-consistency trade-off. While some applications adopt a strong or eventual consistency model, there are others that fall between the two extremes. Users may wish to relax consistency constraints in exchange for better system throughput. In the video recommendation example, observing a newly uploaded video after several seconds is acceptable but keeping users waiting for recommendation harms user experience. Thus, the application can configure the allowed maximal delay for the video updates in order to improve system throughput.

Fourth, vector databases have more stringent and diversified hardware requirements compared with traditional databases. This is attributed to three reasons. First, vector database operations are computation-intensive, and thus hardware accelerators such as GPUs are critical for computing functionalities such as search and indexing. Second, accesses to vector data (e.g., search or update) generally have poor locality, thereby requiring large RAM for good performance. Third, different applications vary significantly in their resource demands for the system functionalities.    
Core functionalities of a vector database include data insertion, indexing, filtering, and vector search. Applications such as video recommendation require online insertion and high concurrency vector search. In contrast, for interactive use cases such as drug discovery, offline data ingestion and indexing are generally acceptable. Although interactive applications usually require lower throughput than recommendation systems, they have high demands for real-time filtering, similarity-based vector search, and hybrid queries. The high hardware costs as well as diverse workload features call for fine-grained elasticity.

The key design goals of \name~are summarized below; these design goals not only fully encompass the above characteristics but also share some common goals with generic cloud-based databases.

\squishlist

\item \textbf{Long-term evolvability:} Overall system complexity must be controlled for the continuous evolution of \name's functionalities. Without the need to support complex transactions, there lies an opportunity to model all the event sequences (such as WAL and inter-component messages) as message queues to cleanly decouple the entire system. In this way, individual components can evolve, be added, or be replaced easily with minimal interference to other components. This design echos large-scale data analytic platforms, which often rely on data streaming systems such as Kafka to connect system components.

\item \textbf{Tunable consistency:} To enable flexible consistency-performance trade-off, \name~should introduce \textit{delta consistency} that falls between strong consistency and eventual consistency, where a read operation returns the last value that was produced at most delta time units preceding itself. It's worth noting that strong consistency and eventual consistency can be realized as special cases of this model, with delta being zero and infinity, respectively.



\item \textbf{Good elasticity:} Workload fluctuations can cause different loads on individual system components. In order to dynamically allocate compute resources to high-load tasks, components must be carefully decoupled, taking both functionality and hardware dependencies into consideration. System elasticity and resource isolation should be managed at the component-level rather than at the system-level (e.g. decoupling indexing from querying versus decoupling read from write).

\item \textbf{High availability:} Availability is a must-have for modern cloud-based applications; \name~must isolate system failures at the component level and make failure recovery transparent.


\item \textbf{High performance:} Query processing performance is key to vector databases. For good performance, implementations to be extensively optimized for hardware. Moreover, the framework should be carefully designed so as to minimize system overheads for query serving. 


\item \textbf{Strong adaptability:} Our customers use vector databases in a variety of environments, ranging from prototyping on laptops to large-scale deployments on the cloud. A vector database should provide consistent user experience and reduce code/data migration overhead across environments.

\squishend

\section{The Manu System} \label{sec:arch}

In this section, we begin by first introducing the basic concepts of \name. Next, we present the system designs, including the overall system architecture, the log backbone, and how \name~conducts vector searches and builds vector search indexes.

\subsection{Schema, Collection, Shard, and Segment} 
\begin{figure}[!t]	
	\centering 
	\includegraphics[width=1\linewidth]{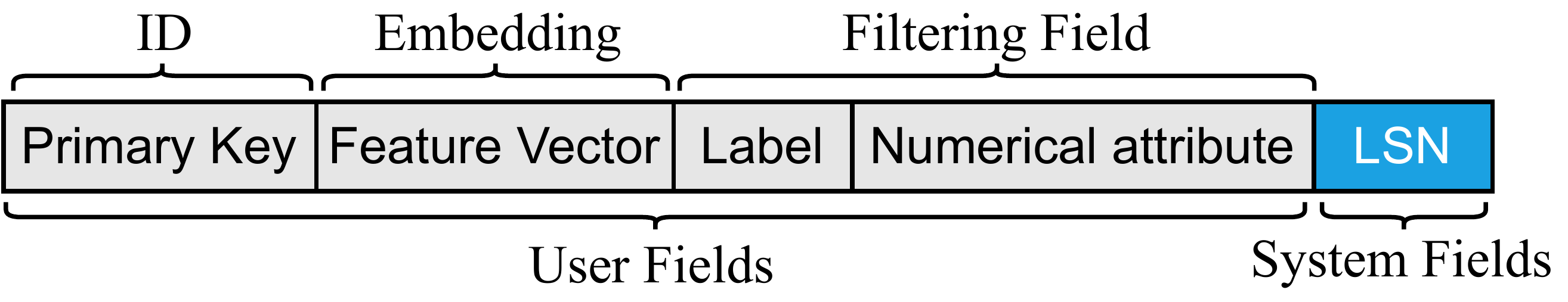}
	\vspace{-6mm}
	\caption{An example of \name's schema.}
	\vspace{-3mm} 
	\label{fig:schema}
	\vspace{-3mm} 
\end{figure}

\stitle{Schema} 
The basic data types of \name~are vector, string, boolean, integer, and floating point. A schema example is given in Figure~\ref{fig:schema}. Suppose each entity consists of five fields and corresponds to a product on an e-commerce platform. The \textit{Primary key} is the ID of the entity. It can either be an integer or a string. If users do not specify this field, the system will automatically add an integer primary key for each entity. The \textit{Feature vector} is the embedding of the product. The \textit{Label} is the category of the product, such as food, book, and cloth. The \textit{Numerical attribute} is a float or an integer associated with the product, such as price, weight, or production date. \name~supports multiple labels and numerical attributes in each entity. Note that these fields are used for filtering, rather than join or aggregation. The \textit{Logical sequence number} (LSN) is a system field hidden from users.

\stitle{Collection}
A Collection is a set of entities similar to the concept of tables in relational databases. For example, a collection can contain all the products of an e-commerce platform. The key difference is that collections have no relations with each other; thus, relational algebra, such as join operations, are not supported.

\stitle{Shard}
The Shard correspondence to insertion/deletion channel. Entities are hashed into multiple shards based on their primary keys during insertion/deletion. \name's data placement unit is segment rather than shard.\footnote{Using segments for data placement is more flexible than shards, as the number of shards is static, while the number of segments grows as the volume of the collection increases.}


\stitle{Segment}
Entities from each shard are organized into segments. A segment can be in either a \textit{growing} or \textit{sealed} state. Sealed segments are read-only while growing segments can accept new entities. A growing segment will switch to sealed state when it reaches a predefined size (set to 512MB by default) or if a period of time has passed without an insertion (e.g., 10 seconds). As some segments may be small (e.g., when insertion has a low arrival rate), \name~merges small segments into larger ones for search efficiency.

\begin{figure}[!t]	
	\centering 
	\includegraphics[width=0.9\linewidth]{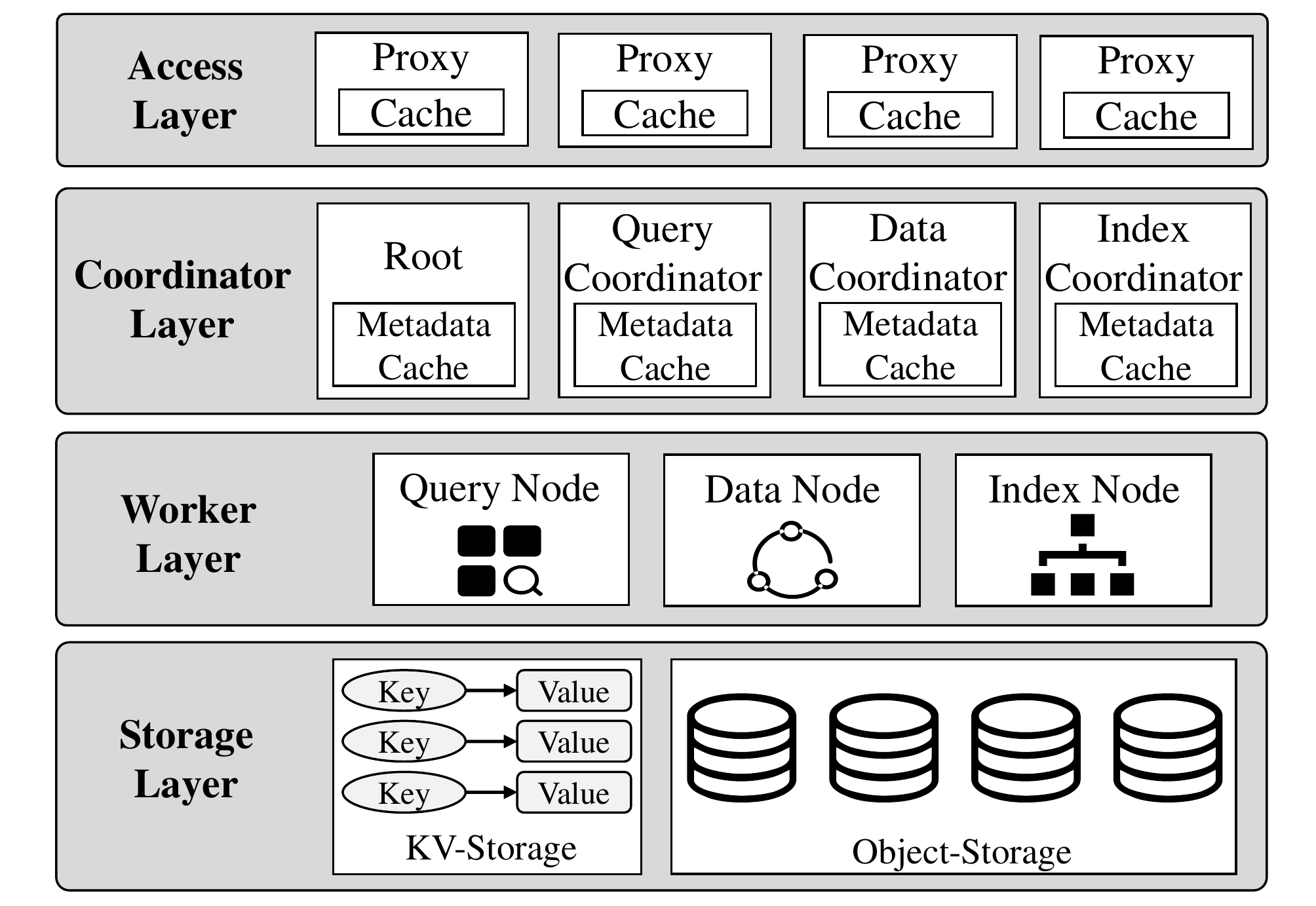}
	\vspace{-3mm}
	\caption{The Architecture of \name.}
	\vspace{-2mm} 
	\label{fig:arch}
	\vspace{-2mm}
\end{figure}

\subsection{System Architecture}


\name~adopts a service-oriented design~\cite{papazoglou2006service} to achieve fine-grained decoupling among the system components. As shown in Figure~\ref{fig:arch}, from top to bottom, \name~has four layers, i.e., \textit{access layer}, \textit{coordinator layer}, \textit{worker layer}, and \textit{storage layer}.

\sntitle{Access layer} consists of stateless proxies that serve as the user endpoints. They work in parallel to receive requests from clients, distribute the requests to the corresponding processing components, and aggregate partial search results before returning to clients. Furthermore, the proxies cache a copy of the metadata for verifying the legitimacy of search requests (e.g., whether the collection to search exists). Search request verification is lightweight and moving it to the proxies has two key benefits. First, requests that fail verification are rejected early, thus lowering the load on other systems components.
Second, it reduces the number of routing hops for requests, thus shortening  request processing latency. 



\sntitle{Coordinator layer} manages system status, maintains metadata of the collections, and coordinates the system components for processing tasks. There are four coordinators, each responsible for different tasks. \textit{Root coordinator} handles data definition requests, such as creating/deleting collections, and maintains meta-information of the collections. \textit{Data coordinator} records detailed information about the collections (e.g., the routes of the segments on storage), and coordinates the data nodes to transform data update requests into binlogs~\cite{binlog}. \textit{Query coordinator} manages the status of the query nodes, and adjusts the assignment of segments (along with related indexes) to query nodes for load balancing. \textit{Index coordinator} maintains meta-information of the indexes (e.g., index types and storage routes), and coordinates index nodes in index building tasks. A coordinator can have multiple instances (e.g., one main and two backups) for reliability. 
As vector databases usually do not have the cross table operations that relational databases have, different collections can be served by separate coordinator instances for throughput. 


\sntitle{Worker layer} conducts the actual computation tasks. The worker nodes are stateless---they fetch read-only copies of data to conduct tasks and do not need to coordinate with each other. This ensures that computation intensive (thus expensive) worker nodes can be easily scaled on demand. We use different worker nodes for different tasks, i.e., \textit{query nodes} for query processing, \textit{index nodes} for index building, and \textit{data nodes} for log archiving. Due to the fact that the workloads for different computation tasks vary significantly over time and across applications, each worker type can scale independently. This design also achieves resource isolation as different computation tasks have different QoS requirements.

\sntitle{Storage layer} persists system status, metadata, the collections, and associated indexes. \name~uses etcd~\cite{ref:etcd} (a key-value store) to host system status and metadata for the coordinators as etcd provides high availability with its leader election mechanism for failure recovery. When metadata is updated, the updated data is first written to etcd, and then synchronized to coordinators. 
Since the volume of other data (e.g., binlog, data, index) is large, \name~ uses AWS S3~\cite{s3} (an object storage) for persistence due to its high availability and low cost.
The API of many other object storage systems is compatible with AWS S3. This allows \name~to easily swap storage engines, if necessary. At present, storage engines including AWS S3, MinIO~\cite{minio}, and Linux file system are supported.
Note that the high latency that comes with object storage is not a performance bottleneck as the worker nodes conduct computation tasks on in-memory, read-only copies of data. 




\subsection{The Log Backbone}


\begin{figure}[!t]	
	\centering 
	\includegraphics[width=1\linewidth]{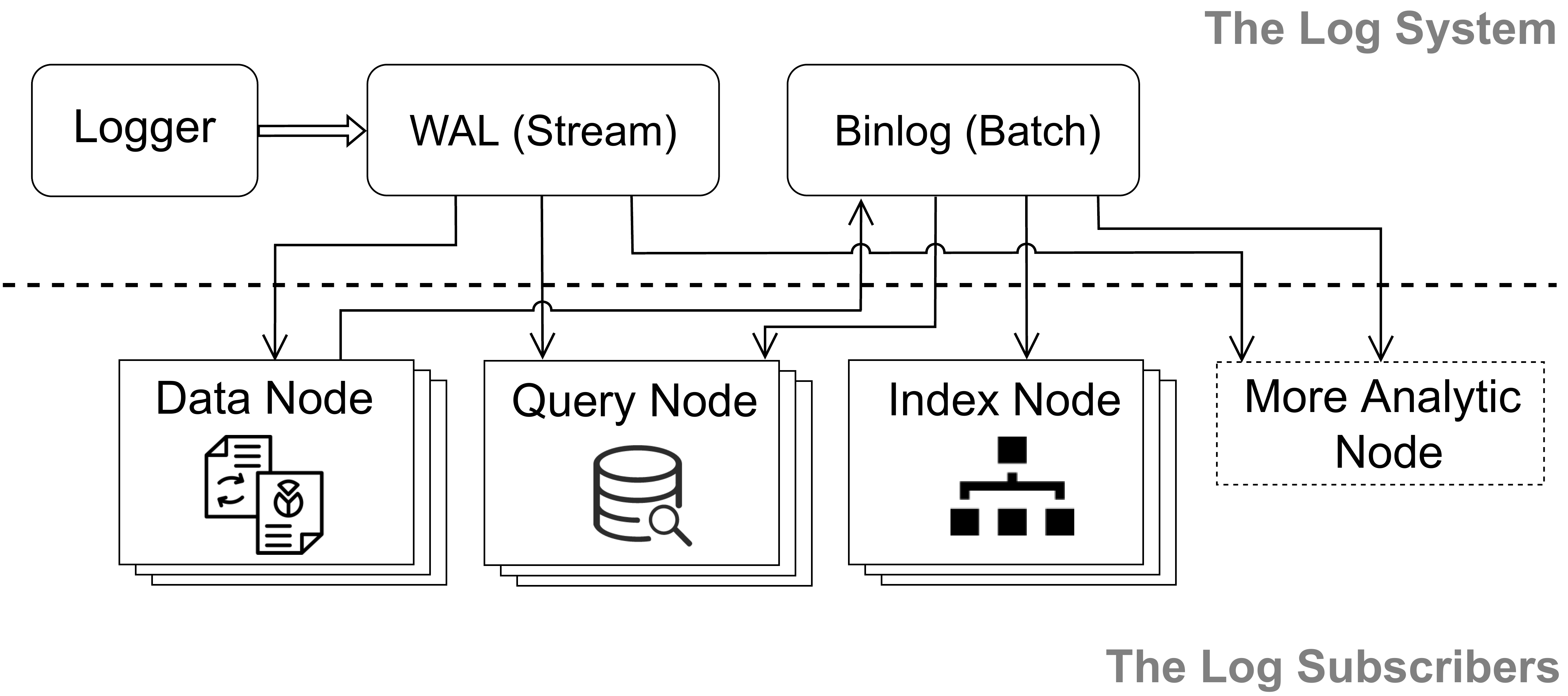}
	\vspace{-4mm}
	\caption{Overview of \name's log system.}
	\vspace{-3mm}
	\label{fig:log-overall}
	\vspace{-3mm}
\end{figure}


The log system is the backbone of \name, which connects the decoupled system components. As shown in Figure~\ref{fig:log-overall}, \name~exposes the write-ahead log (WAL) and binlog as backbone services. The WAL is the incremental part of system log while the binlog is the base part; they complement each other in delay, capacity and cost. 
Loggers are entry points for publishing data onto the WAL.
Data nodes subscribe to the WAL and convert the row-based WALs into column-based binlogs. 
All read-only components such as index nodes and query nodes are independent subscribers to the log service to keep themselves up-to-date.
This architecture completely decouples the write and read components, thus allowing the components (e.g., WAL, binlog, data nodes, index nodes and query nodes) to scale independently.

\name~ records all the requests that change system state to the log, including data definition requests (e.g., create/delete collection), data manipulation requests (e.g., insert/delete a vector), and system coordination messages (e.g., load/dump a collection to/from memory). Note that vector search requests are not written to the log as they are read-only operations and do not change system state. We use logical logs instead of physical logs, as logical logs focus on event recording, rather than describing the modifications to physical data pages. This allows the subscribers to consume the log data in different ways depending on their functions.

\begin{figure}[!t]	
	\centering 
	\includegraphics[width=1\linewidth]{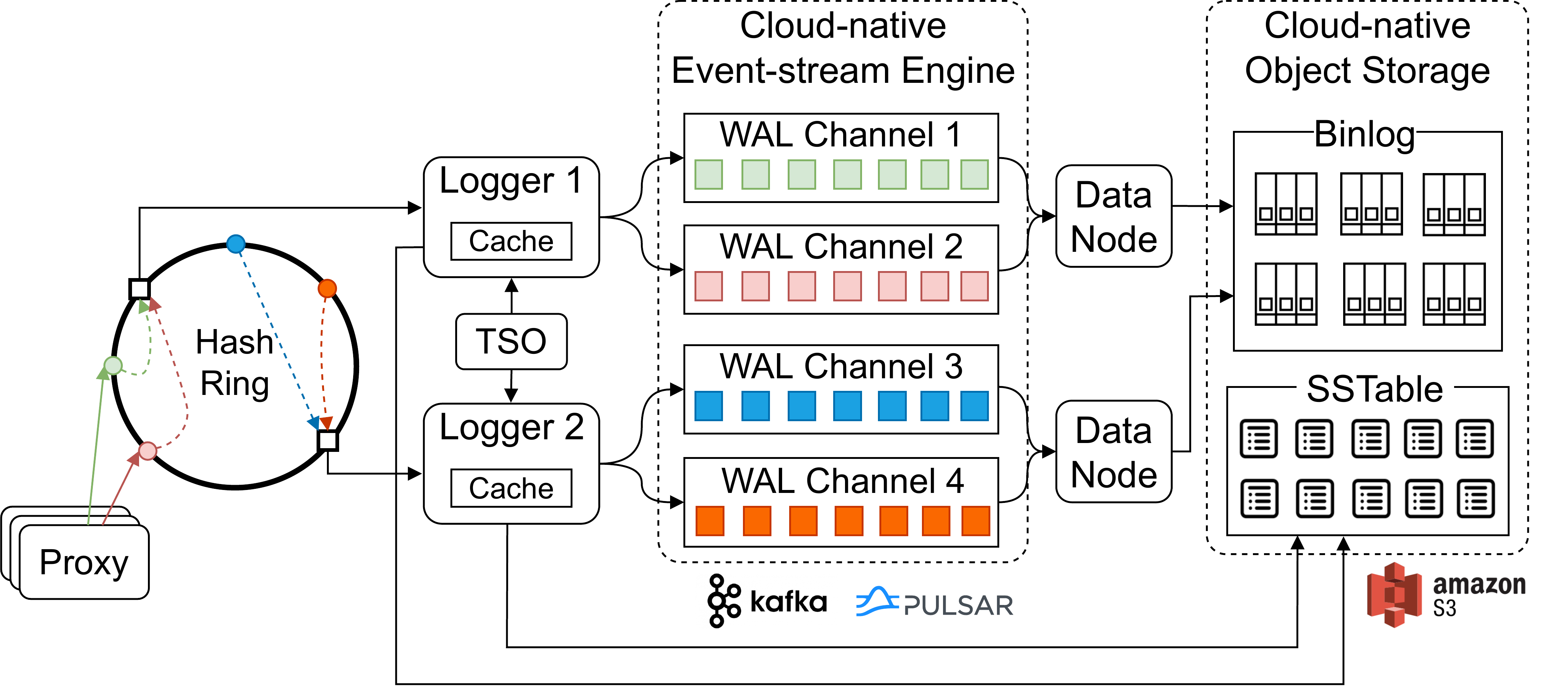}
	\vspace{-1mm}
	\caption{Detailed structure of \name's log system.}
	\vspace{-1mm}
	\label{fig:log-detail}
\end{figure}

Figure~\ref{fig:log-detail} illustrates the detailed architecture of the log system. For the sake of clarity, we only illustrate the parts related to insert requests. 
The loggers are organized in a hash ring, and each logger handles one or more logical buckets in the hash ring based on consistent hashing.
Each shard corresponds to a logical bucket in the hash ring and a WAL channel.
Each entity in insert requests is hashed to a shard (and thus channel) based on their ID.
When a logger receives a request, it will first verify the legibility of the request, assign an LSN for the logged entity by consulting the central time service oracle (TSO), determine the segment the entity should go to, and write the entity to WAL. The logger also writes the mapping of the new entity ID to segment ID into a local LSM tree and periodically flushes the incremental part of the LSM tree to object storage, which keeps the entity to segment mapping using the SSTable format of RocksDB. Each logger caches the segment mapping (e.g., for checking if the entity to delete exists) for the shards it manages by consulting the SSTable in object storage.

The WAL is row-based and read in a streaming manner for low delay and fine-grained log pub/sub. It is implemented via a cloud-based message queue such as Kafka or Pulsar. We use multiple logical channels for the WAL in order to prevent different types of requests from interfering with each other, thus achieving a high throughput. Data definition requests and system coordination messages use their own channels while data manipulation requests hashed across multiple channels to increase throughput.

Data nodes subscribe to the WAL and convert the row-based WALs into column-based binlogs. Specifically, values from the same field (e.g., attribute and vector) from the WALs are stored together in a column format in binlog files.
The column-based nature of binlog makes it suitable for reading per field values in batches, thus increasing storage and IO efficiency. 
An example of this efficiency comes with the index nodes. Index nodes only read the required fields (e.g., attribute or vector) from the binlog for index building and thus are free from the read amplifications.

\stitle{System coordination} Inter-component messages are also passed via log, e.g., data nodes announce when segments are written to storage and index nodes announce when indexes have been built. This is because the log system provides a simple and reliable mechanism for broadcasting system events. Moreover, the time semantics of the log system provide a deterministic order for coordination messages. For example, when a collection should be released from memory, the query coordinator publishes the request to log, and does not need to confirm whether the query nodes receive the message or handle query node failure. The query nodes independently subscribe to the log and asynchronously release segments of the collection.

\subsection{Tunable Consistency}

We adopt a delta consistency model to enable flexible performance-consistency trade-offs, which guarantees a bounded staleness of data seen by search queries. Specifically, the data seen by a query can be stale for up to delta time units, with respect to time of the last data update, where delta is an user-specified “staleness tolerance” given in virtual time. 

In practice, users prefer to define temporal tolerance as physical time, e.g., 10 seconds. \name~achieves this by making the LSN assigned to each request extremely close to physical time.
\name~uses a hybrid logical clock in the TSO to generate timestamps. Each timestamp has two components: a physical component that tracks physical time, and a logical component that tracks event order. The logical component is needed since multiple events may happen at the same physical time unit.
Since a timestamp is used as a request's LSN, the value of the physical component indicates the physical time when the request was received by \name.

For a log subscriber, e.g., a query node, to run the delta consistency model, it needs to know three things: (1) the user-specified staleness tolerance $\tau$, (2) the time of the last data update, and (3) the issue time of the search request. 
In order to let each log subscriber know (2), we introduce a time-tick mechanism. Special control messages called time-ticks (similar to watermarks in Apache Flink~\cite{flink}) are periodically inserted into each log channel (for example, WAL channel) signaling the progress of data synchronization. Denote the latest time-tick a subscriber consumed as $L_s$ and the issue time of a query as $L_r$, if $L_r - L_s < \tau$ is not satisfied, the query node will wait for the next time-tick before executing the query.

Note that strong consistency and eventual consistency are two special cases of delta consistency, where delta equals to 0 and infinity, respectively. To the best of our knowledge, our work is the first to support delta consistency in a vector database.

\subsection{Index Building} 

\begin{table}[t]
	\centering
	\caption{Major indexes in \name}
	\vspace{-2mm}
	\label{tab:index}
	\begin{center}
		\fontsize{8}{9}\selectfont
		\begin{tabular}{|c|c|}
			\hline
			\textbf{Vector Quantization}  & PQ, OPQ, RQ, SQ \\
			\hline
			\textbf{Inverted Index} & IVF-Flat, IVF-PQ, IVF-SQ, IVF-HNSW, IMI\\
			\hline
			\textbf{Proximity Graph} & HNSW, NSG, NGT\\
			\hline
			\textbf{Numerical Attribute} & B-Tree, Sorted List \\
			\hline
		\end{tabular}
	\end{center}
	\vspace{-2mm}
\end{table}


Searching similar vectors in large collections by brute-force, i.e., scanning the whole dataset, usually yields unacceptably long delays.
Numerous indexes have been proposed to accelerate vector search and \name~automatically builds user specified indexes. Table~\ref{tab:index} summarizes the indexes currently supported by \name, and we are continuously adding new indexes following the latest indexing algorithms. These indexes differ in their properties and use cases.
Vector quantization (VQ)~\cite{jegou2010product, ge2013optimized} methods compress vectors to reduce memory footprint and the costs for vector distance/similarity computation.
For example, scalar quantization (SQ)~\cite{zhou2012scalar} maps each dimension of vector (data types typically are int32 and float) to a single byte. 
Inverted indexes~\cite{scholer2002compression} group vectors into clusters, and only scan the most promising clusters for a query.
Proximity graphs~\cite{malkov2018efficient,fu2019fast,iwasaki2018optimization} connect similar vectors to form a graph, and achieve high accuracy and low latency at the cost of high memory consumption~\cite{li2019approximate}.
Besides vector indexes, \name~also supports indexes on the attribute field of the entities to accelerate attribute-based filtering.

There are two index building scenarios in \name, i.e., \textit{batch indexing} and \textit{stream indexing}.
Batch indexing occurs when the user builds  an index for an entire collection (e.g., when all vectors are updated with a new embedding model). In this case, the index coordinator obtains the paths of all segments in the collection from the data coordinator, and instructs index nodes to build indexes for each segment. Stream indexing happens when users continuously insert new entities, and indexes are built asynchronously on-the-fly without stopping search services. Specifically, after a segment accumulates a sufficient number of vectors, its resident data node seals the segment and writes it to object storage as a binlog. The data coordinator then notifies the index coordinator, which instructs a index node to build index for the segment. The index node loads only the required column (e.g., vector or attribute) of the segment from object storage for indexing building to avoid read amplification. For entity deletions, \name~uses a bitmap to record the deleted vectors and rebuilds the index for a segment when a sufficient number of its entities have been deleted. In both batch and stream indexing scenarios, after the required index is built for a segment, the index node persists it in the object storage and sends the path to the index coordinator, which notifies the query coordinator so that query nodes can load the index for processing queries. The index coordinator also monitors the status of the index nodes and  shuts down idle index nodes to save costs. As vector indexes generally have sub-linear search complexity w.r.t. the number of vectors, searching a large segment is cheaper than several small segments, \name~builds joint indexes on multiple segments when appropriate.

\subsection{Vector Search}

\name~supports \textit{classical vector search}, \textit{attribute filtering}, and \textit{multi-vector search}. For classical vector search, the distance/similarity function can be Euclidean distance, inner product or angular distance. Attribute filtering is useful when searching vectors similar to the query subject to some attribute constraints. For example, an e-commerce platform may want to find products that interest the customer and cost less than 100\$. \name~supports three strategies for attribute filtering and uses a cost-based model to choose the most suitable strategy for each segment. Multi-vector search is required when an entity is encoded by multiple vectors, for example, a product can be described by both embeddings of its image and embeddings of its text description. In this case, the similarity function between entities is defined as a composition of similarity functions on the constituting vectors. \name~supports two strategies for multi-vector search and chooses the one to use according to the entity similarity function. For more details about how \name~handles attribute filtering and multi-vector search, interested readers can refer to Milvus~\cite{wang2021milvus}.

For vector search, \name~partitions a collection into segments and distributes the segments among query nodes for parallel execution.~\footnote{\name~loads all data to the query nodes as different queries may access different parts of the data, and a hot compute side cache is necessary for low latency. This is different from general cloud DBMSs that decouple compute and storage (e.g., Snowflake~\cite{DagevilleCZAABC16}), which only fetch the required data to compute side upon request.} The proxies cache a copy of the distribution of segments on query nodes by inquiring the query coordinator, and dispatch search requests to query nodes that hold segments of the searched collection. The query nodes perform vector searches on their local segments without coordination using a \textit{two-phase reduce} procedure. For a top-$k$ vector search request, the query nodes search their local segments to obtain the segment-wise top-$k$ results. These results are merged by each query node to form the node-wise top-$k$ results. Then, the node-wise top-$k$ results are aggregated by the proxy for the global top-$k$ results and returned to the application. To handle the deletion of vectors, the query nodes use a bitmap to record the deleted vectors in each segment and filter the deleted vectors from the segment-wise search results. Users can configure \name~to batch search requests to improve efficiency. In this case, the proxies organize cache search requests if results of the previous batches have not been returned yet. In the cache, requests of the same type (i.e., target the same collection and use the same similarity function) are organized into the one batch and handled by \name~together. \name~also allows maintaining multiple hot replicas of a collection to serve queries for availability and throughput. 


Query nodes obtain data from three sources, i.e., the WAL, the index files, and the binlog. For data in the growing segments, query nodes subscribe to the WAL and conduct searches using brute force scan so that updates can be searched within a short delay. A dilemma for segment size is that larger size yields better search efficiency once the index is built but brute force scan on growing segment is also more costly. 
To tackle this problem, we divide each segment into \textit{slices} (each containing 10,000 vectors by default). New data are inserted into the slices sequentially, and after a slice is full, a light-weight temporary index (e.g., IVF-FLAT) is built for it. Empirically, we observed that the temporary index brings up to 10X speedup for searching growing segments.
When a segment changes from growing state to sealed state, its index will be built by an index node and then stored in object storage. After that, query nodes are notified to load the index and replace the temporary index.

\begin{table*}[t]
	\centering
	\caption{Main commands of Python-based PyManu API}
	\vspace{-2mm}
	\label{tab:api}
	\begin{tabular}{|l|l|}
		\hline
		\multicolumn{1}{|c|}{\textbf{Methods}}                                    & \multicolumn{1}{c|}{\textbf{Description}}                                                                           \\ \hline
		Collection(\textit{name}, \textit{schema})                                & Create collection with name \textit{str} and schema \textit{schema}                                                 \\ \hline
		Collection.insert(\textit{vec})                                           & Insert vector \textit{vec} into collection                                                                      \\ \hline
		Collection.delete(\textit{expr})                                          & Delete vectors satisfying boolean expression \textit{expr} from collection                                                          \\ \hline
		Collection.create\_index(\textit{field}, \textit{params})    & Create index on a \textit{field} of the vectors using parameters \textit{params}    \\ \hline
		Collection.search(\textit{vec}, \textit{params})       & Vector search for \textit{vec} with parameters \textit{params}                              \\ \hline
		Collection.query(\textit{vec}, \textit{params}, \textit{expr})                                           & Vector search for \textit{vec} with boolean expression \textit{expr} as filters 
		\\ \hline
	\end{tabular}
\end{table*}

Query nodes access the binlog for data when the distribution of segments among the query nodes changes, which may happen during scaling, load-balancing, query node failure and recovery. Specifically, the query coordinator manages the segment distribution and monitors the query nodes for liveness and workload to coordinate failure recovery and scaling.     On failure recovery, the segments and their corresponding indexes (if they exist) handled by failed query nodes are loaded to the healthy ones.~\footnote{The WAL channels subscribed to by failed query nodes are also assigned to healthy ones.}
In the case of scaling down, a query node can be removed once other query nodes load the indexes for the segments it handles from the object storage.
When scaling up, the query coordinator assigns some of the segments to the newly added nodes. A new query node can join after it loads the assigned segments, and existing query nodes can release the segments no longer handled by them. The query coordinator also balances the workloads (and memory consumption) of the query nodes by migrating segments. Note that \name~does not ensure that segment redistribution is atomic, and a segment can reside on more than one query node. This does not affect correctness as the proxies remove duplicate result vectors for a query.

\section{Feature Highlights} \label{sec:feature}

In this part, we introduce several key features of \name~for usability and performance.

\subsection{Cloud Native and Adaptive}

The primary design goal of \name~is to be a cloud native vector database such that it fits well into cloud-based data pipelines. To this end, \name~decouples system functionalities into storage, coordinators, and workers in the overall design. For storage, \name~uses a transaction KV for metadata, message queues for logs, and an object KV for data, which are all general storage services provided by major cloud vendors and thus enables easy deployment. For coordinators that manage system functionalities, \name~uses the standard one main plus two hot backups configuration for high availability. For workers, \name~decouples vector search, log archiving and index building tasks for component-wise scaling, a model suitable for cloud-based on-demand resource provisioning. The log backbone allows the system components to interact by writing/reading logs in their own ways. This enables the system components to evolve independently and makes it easy to add new components. The log backbone also provides consistent time semantics in the system, which are crucial for deterministic execution and failure recovery.

Our customers use vector databases in the entire life-cycle of their applications. For example, an application usually starts with data scientists conducting proof of concept (PoC) on their personal computers. Then, it is migrated to dedicated clusters for testing and finally deployed on the cloud. Thus, to reduce migration costs, our customers expect vector databases to adapt to different deployment scenarios while providing a consistent set of APIs. To this end, \name~defines unified interface for the system components but provides different invocation methods and implementations for different platforms. For example, on cloud, local cluster and personal computer, \name~uses cloud service APIs, remote procedure call (RPC) and direct function calls to invoke system functionalities, respectively. The object KV can be the local file system (e.g., MinIO~\cite{minio}) on personal computers, and S3 on AWS. Thus, \name{} applications can migrate with little or no change across different deployment scenarios.

\subsection{Good Usability}




Data pipelines interact with \name{} in simple ways: vector collections, updates for vector data and search requests are fed to \name, and \name{} returns the identifiers of the search results for each search request, which can be used to retrieve objects (e.g.. images, advertisements, movies) in other systems. Because different users adopt different programming languages and development environments, \name~provides APIs in popular languages including Python, Java, Go, C++, along with RESTful APIs. As an example, we show key commands of the Python-based Py\name~API in Table~\ref{tab:api}, which uses the object-relational mapping (ORM) model and most commands are related to the \textit{collection} class. As shown in Table~\ref{tab:api}, Py\name~allows users to manage collections and indexes, update collections, and conduct vector searches. The \textit{search} command is used for similarity-based vector search while the \textit{query} command is mainly used for attribute filtering. We show an example of conducting top-$k$ vector search by specifying the parameters in \textit{params} in as follows. 
\begin{Verbatim}[fontsize=\small,label=code]
    query_param = {
        "vec": [[0.6, 0.3, ..., 0.8]],
        "field": "vector",
        "param": {"metric_type": "Euclidean"},
        "limit": 2,
        "expr": "product_count > 0",
    }
    res = collection.search(**query_param)
\end{Verbatim}
In the above example, the search request provides a high dimensional vector $[0.6, 0.3, ..., 0.8]$ as query and searches the \textit{feature vector} field of the collection. The similarity function is Euclidean distance and the targets are the top-2 similar vectors in the collection (i.e., with $\mathsf{limit}=2$). 

\begin{figure}[!t]	
	\centering 
	\includegraphics[width=0.9\linewidth]{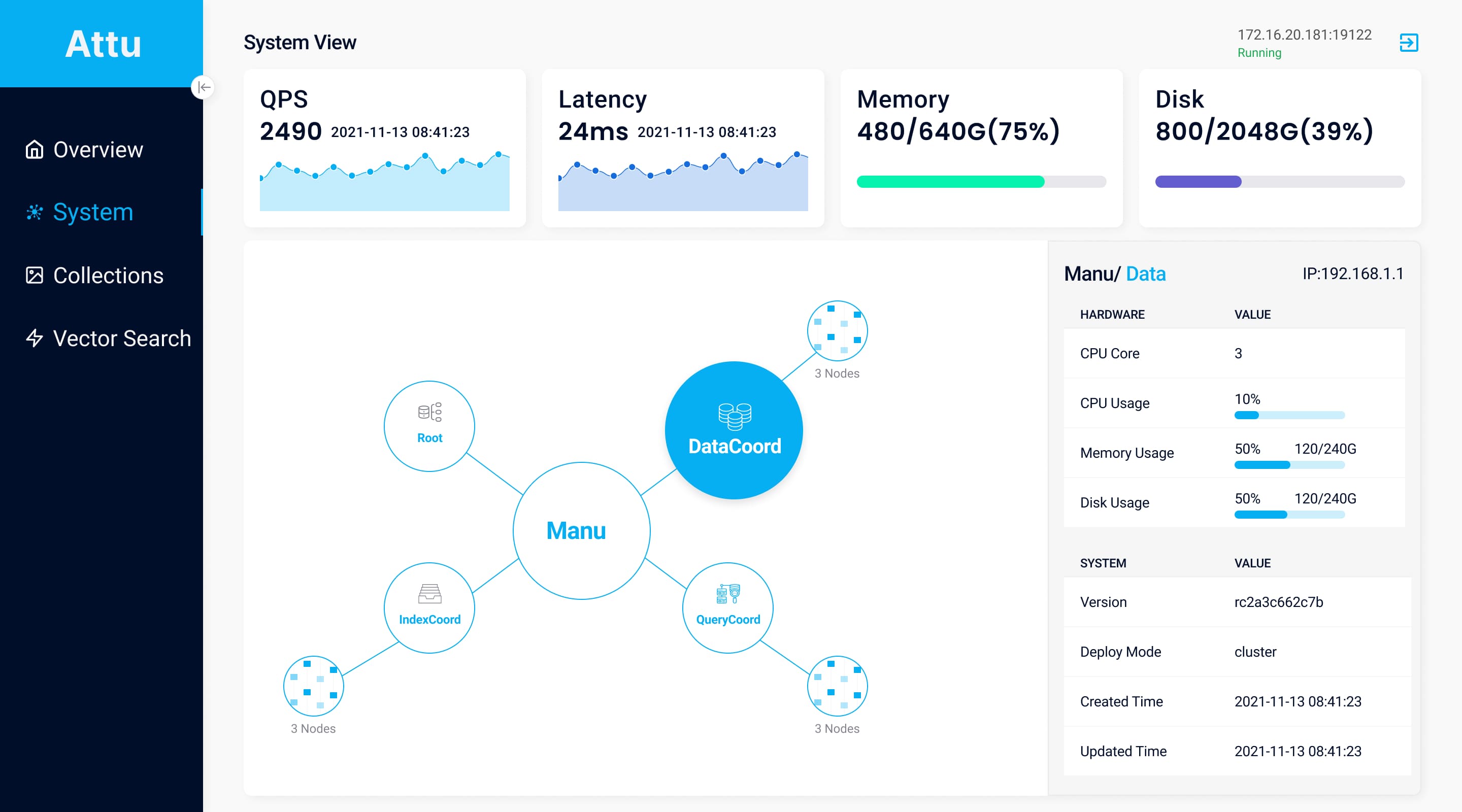}
	\vspace{-2mm}
	\caption{A screenshot of Attu, the GUI tool of \name{}.}
	\vspace{-3mm}
	\label{fig:attu}
	\vspace{-2mm}
\end{figure}

For easy system management, \name~provides a GUI tool called Attu, for which a screen shot is shown in Figure~\ref{fig:attu}. In the \textit{system view}, users can observe overall system status including queries processed per second (QPS), average query latency, and memory consumption on the top of screen. By clicking a specific service (e.g., data service), users can view detailed information of the worker nodes for the service on the side. We also allow users to add and drop worker nodes with mouse clicks. In the \textit{collection view}, users can check the collections in the system, load/dump collections to/from memory, delete/import collections, check the index built for the collections, and build new indexes. In the \textit{vector search view}, users can check the search traffic and performance on each collection, configure the index and search parameters to use for each collection. The vector search view also allows to issue queries for functionality test.

For vector search, using different parameters for the indexes (e.g., neighbor size $M$ and queue size $L$ for HNSW~\cite{malkov2018efficient}) yields different trade-offs among cost, accuracy, and performance. However, even experts find it difficult to set proper index parameters as the parameters are interdependent and their influences vary across collections. \name~adopts a Bayesian Optimization with Hyperband (BOHB)~\cite{falkner2017combining} method to automatically explore good index parameter configurations. Users provide a utility function to score the configurations (e.g., according to search recall, query throughput) and set a budget to limit the costs of parameter search. BOHB starts with a group of initial configurations and evaluates their utilities. Then, Bayesian Optimization is used to generate new candidate configurations according to historical trials and Hyperband is used to allocate budgets to different areas in the configuration space. The idea is to prioritize the exploration of areas close to high utility configurations to find even better configurations. \name~also supports sampling a subset of the collection for the trails to reduce search costs. We are still improving the automatic parameter search module and plan to extend it to searching system configurations (e.g., the number and type of query nodes).


\subsection{Time Travel}

Users often need to rollback the database to fix corrupted data or code bugs. \name~allows users to specify a target physical time $T$ for database restore, and jointly uses checkpoint and log replay for rollback. We mark each segment with its progress $L$ and periodically checkpoints the segment map for a collection, which contains information (such a route, rather than data) of all its segments. To restore the database at time $T$, we read the closest checkpoint before $T$, load all segments in the segment map and replay the WAL log for each segment from its local progress $L$. This design reduces storage consumption as we do not write entire collection for each checkpoint. Instead, segments that have no changes are shared among checkpoints. The replay overhead is also reduced as each segment has its own progress. Users can also specify a expiration period to delete outdated log and segments to reduce storage consumption.

\subsection{Hardware Optimizations}
\name~comes with extensively optimized implementations for CPU, GPU and SSD for efficiency. For more details about our CPU and GPU optimizations, interested readers can refer to Milvus~\cite{wang2021milvus}.


SSD is 100x cheaper than dram and offers 10x larger bandwidth than HDD. thus, \name~supports using SSD to store large vector collections on cheap query nodes with limited dram capacity. the challenge is that SSD bandwidth is still much smaller than dram, which may lead to low query processing throughput and thus necessitates careful designs for storage layout and index structure. as SSD reads are conducted with  4kb blocks (i.e., reading less than 4kb has the same cost as reading 4kb), \name~organizes the vectors into buckets whose sizes are close to but smaller than 4kb.~\footnote{we set the bucket size to a few times (e.g., 4 and 8) of 4kb if the size of an individual vector is large.} this is achieved by conducting hierarchical k-means for the vectors and controlling the sizes of the clusters. each bucket is stored on 4kb aligned blocks on SSD for efficient read and represented by its k-means center in dram. these centers are organized using existing vector search indexes (e.g., ivf-flat, hnsw).

vector search with SSD is conducted in two stages. first, we search the cluster centers in dram for the ones that are most similar to the query. then, the corresponding buckets are loaded from SSD for scan. to reduce the amount of data fetched from SSD, we compress the vectors using scalar quantization, which has negligible influence on the quality of search results according to our trials. another problem is that k-means can put vectors similar to a query into several buckets but the centers of some buckets may not be similar to the query, which leads to a low recall. to tackle this problem, \name~uses a strategy similar to multiple hash tables in locality sensitive hashing~\cite{indyk1998approximate}. hierarchical k-means is conducted by multiple times, each time assigning a vector to a bucket. this means that a vector is replicated multiple times in SSD and we index all cluster centers for bucket search in dram. \name's SSD solution wins track 2 (search with SSD) of the billion-scale approximate nearest neighbor search challenge at neurips'2021~\cite{challenge}. tests results show that \name's solution improves the recall of the competition baseline by up to 60\% at the same query processing throughput.\footnote{for more details about results please refer to ~\cite{simhadri2022results}.} we notice that another work adopts similar designs for SSD-based vector search~\cite{spann}.

\section{Use Cases and Evaluation} \label{sec:case}


\begin{figure}[!t]	
	\centering 
	\includegraphics[width=0.35\textwidth]{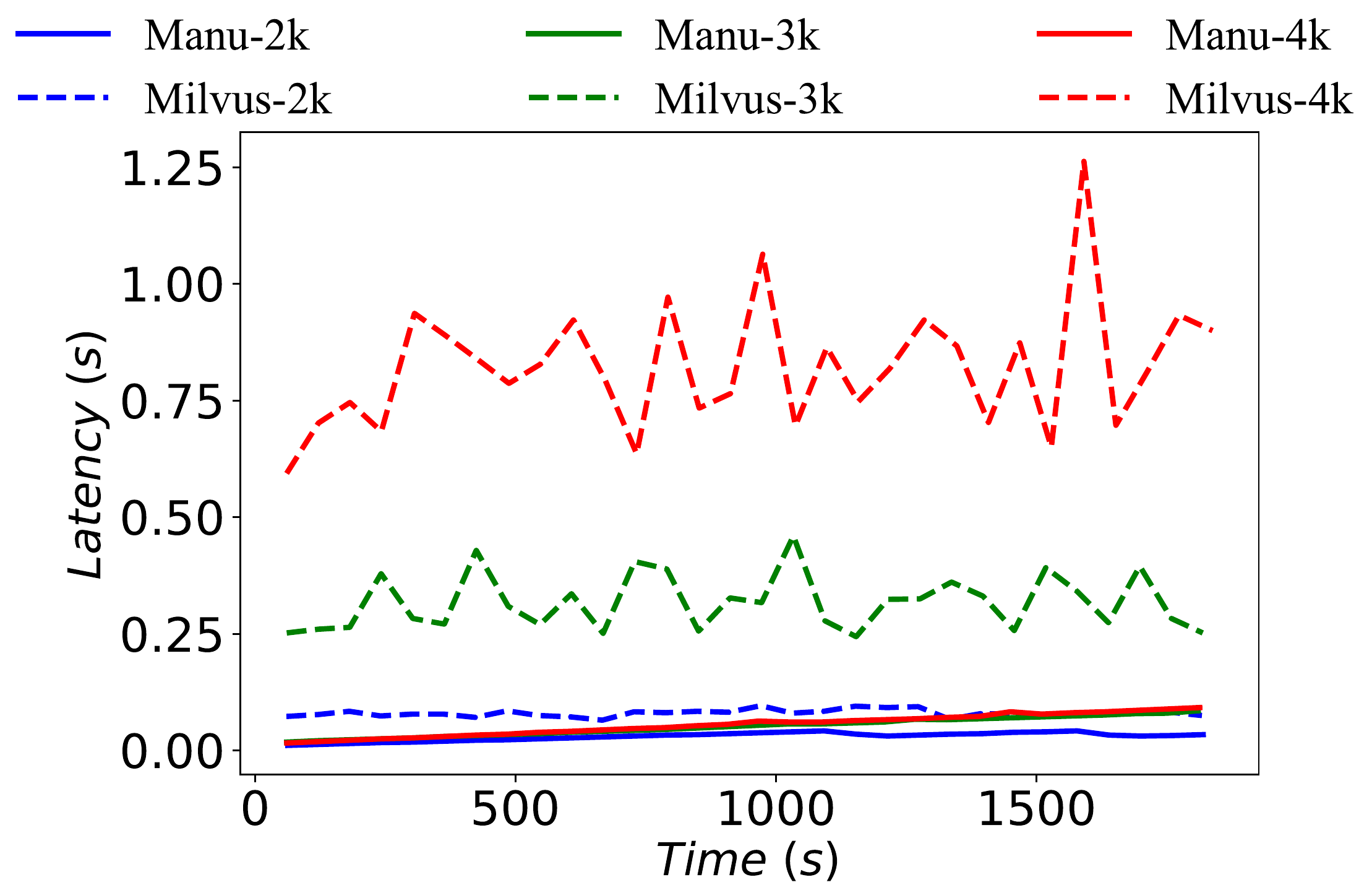}
	\vspace{-2mm}
	\caption{\name{} and Milvus for mixed workloads, numbers behind legends (e.g., 1k) indicate insertion rate.}
    \vspace{-2mm}
	\label{fig:mixed workloads}
    \vspace{-2mm}
\end{figure} 

Before introducing the use cases of \name, we first compare \name{} with Milvus, our previous vector database. Milvus adopts an eventual consistency model and thus does not support the tunable consistency of \name{}. To show the advantages brought by \name{}'s fine-grained functionality decomposition, we create a mixed workload. Specifically, we start with an empty collection, insert vectors at a fixed rate (e.g., 2k vectors per second), and measure the latency for search requests over time. Both \name{} and Milvus use 6 nodes and are properly configured for good performance. The results in Figure~\ref{fig:mixed workloads} show that the search latency of Milvus is significantly longer than \name{}, especially when insertion rates are high (e.g., at 3k and 4k). Milvus has multiple read nodes, but only one write node, to ensure eventual consistency. The write node responsible for data insertion and index construction, and thus write tasks and index building tasks contend for resource. As a result, the index building latency is long and brute force search is used for a large amount of data. In contrast, with dedicated index nodes, \name{} finishes index building quickly and thus search latency remains low over the entire period. 



\begin{figure}[!t]	
	\centering 
	\includegraphics[width=0.9\linewidth]{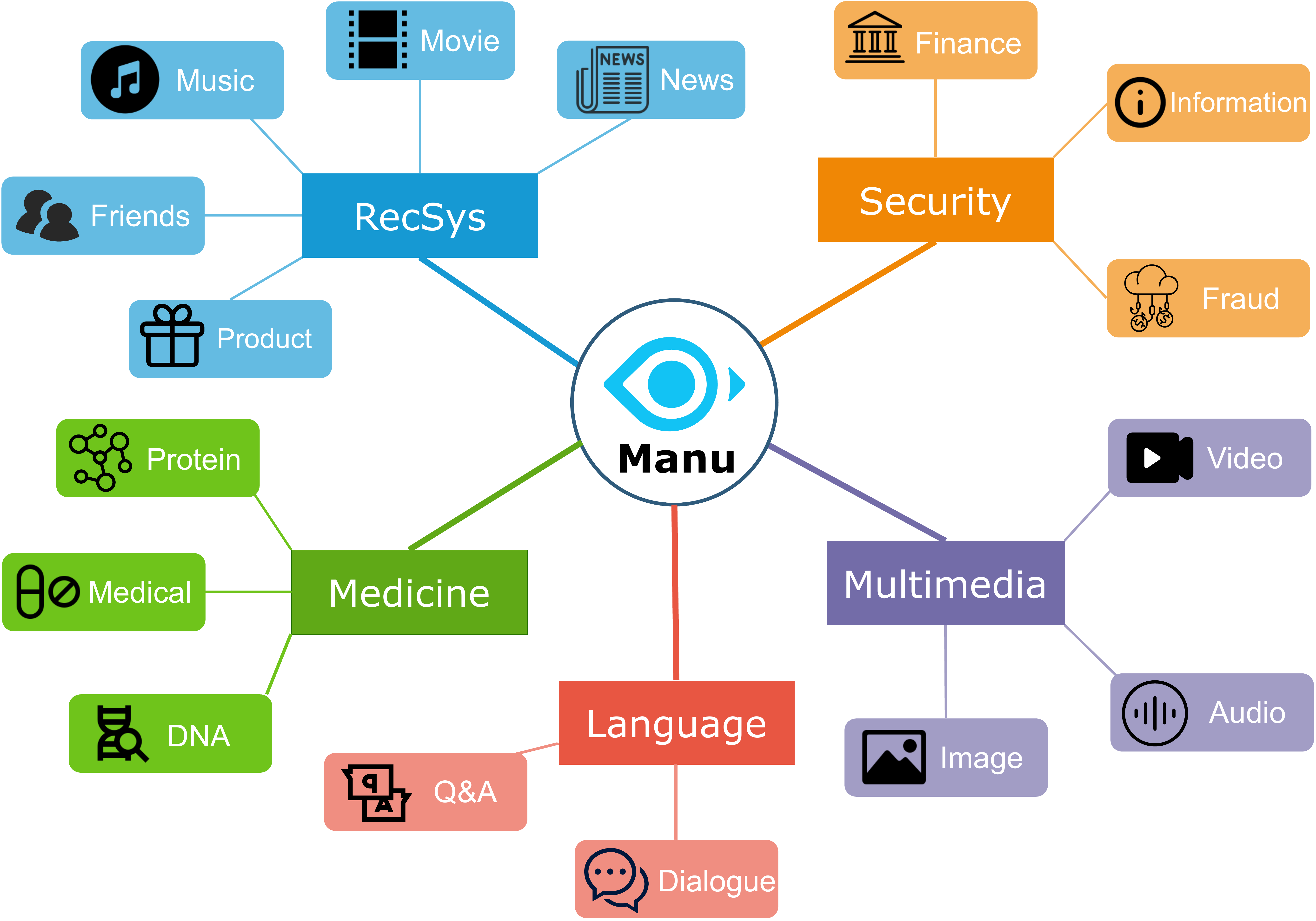}
	\caption{The use cases of \name.}
	\vspace{-3mm}
	\label{fig:application}
	\vspace{-3mm}
\end{figure}



\subsection{Overview of Use Cases}

We classify our customers into 5 application domains in Figure~\ref{fig:application} and briefly elaborate them as follows.

\stitle{Recommendation} Platforms for e-commerce~\cite{wang2018billion}, music~\cite{van2013deep}, news~\cite{liu2010personalized}, video~\cite{covington2016deep}, and social network~\cite{jamali2010matrix} record user-content interactions, and use the data to map users and contents to embedding vectors with techniques such as ALS~\cite{takacs2012alternating} and deep learning~\cite{lecun2015deep}. Finding contents of interest for a user is conducted by searching content vectors having large similarity scores (typically inner product) with user vector.

\stitle{Multimedia} Multimedia contents (e.g., images, video and audio) are becoming increasingly popular, and searches for multimedia contents from large corpus are common online. The general practice is to embed both user query and corpus contents into vectors using tools such as CNN~\cite{lecun1995convolutional} and RNN~\cite{zaremba2014recurrent}. Searching multimedia contents is conducted by finding vectors similar to the user query.

\stitle{Language} Automatic questing answering and machine-based dialogue attract much attention recently with products such as Siri~\cite{siri} and Xiaoice~\cite{xiaoice}, and searches for text contents  is a general need. With models such as Word2Vec~\cite{mikolov2013distributed} and BERT~\cite{bert}, language sequences are embedded into vectors such that retrieving language contents boils down to finding content vectors that are similar to user query. 

\stitle{Security} Blocking spams and scanning viruses are important for security. The common practice is to map spams and viruses into vectors using hashing~\cite{luo2018optimizing} or tailored algorithms~\cite{hersovici1998shark}. After that, suspicious spams and viruses can be checked by finding the most similar candidates in the corpus for further check. 

\stitle{Medicine} Many medical applications search for certain chemical structures and gene sequences for drug discovery or health risk identification. With tools such as GNN~\cite{scarselli2008graph} and LSTM~\cite{xingjian2015convolutional}, chemical structures and gene sequences can be embedded into vectors and their search tasks are cast into vector search.


Full-fledged vector databases are necessary for the forgoing domains as they require much more complex functionality support in addition to vector search. Specifically, as the vector datasets are large and applications have high requirements for throughput, they need distributed computing with multiple nodes for scalability. The vectors are also continuously updated when new user/content comes, user behavior changes or the embedding model is updated. Since most of these applications serve end users, they require high availability and durability. Some of our customers have deployed \name{} in their production environment, and they found \name{} satisfactory in terms of usability, performance, elasticity, and adaptability. In what follows, we simulate some typical application scenarios of our customers to demonstrate the advantages of \name{}.   




\subsection{Example Use Cases}



Due to business security, the names of the customers are anonymous. For the experiments, we use two datasets widely used for vector search research, i.e., SIFT~\cite{sift}(with 128-dim vectors) and DEEP~\cite{deep} (with 96-dim vectors), and extract sub-datasets with the required sizes. By default, we use two query nodes, one data node and one index node for \name. Each worker node is an EC2 m5.4xlarge instance running on Amazon Linux AMI version 5.4.129. For index, we experiment with IVF-FLAT~\cite{johnson2019billion} and HNSW~\cite{malkov2018efficient}, which are widely used in practice. When comparing \name~with other systems, we always ensure that the systems use the same resource and are properly configured. Due to time and expense limits, we are only able to compare with some vector databases in a subset of the experiments. We search the top-50 most similar vectors for each query, and ensure that average search recall is above 0.8 if recall is not reported explicitly. 


\begin{figure}[!t]	
	\centering 
	\includegraphics[width=1\linewidth]{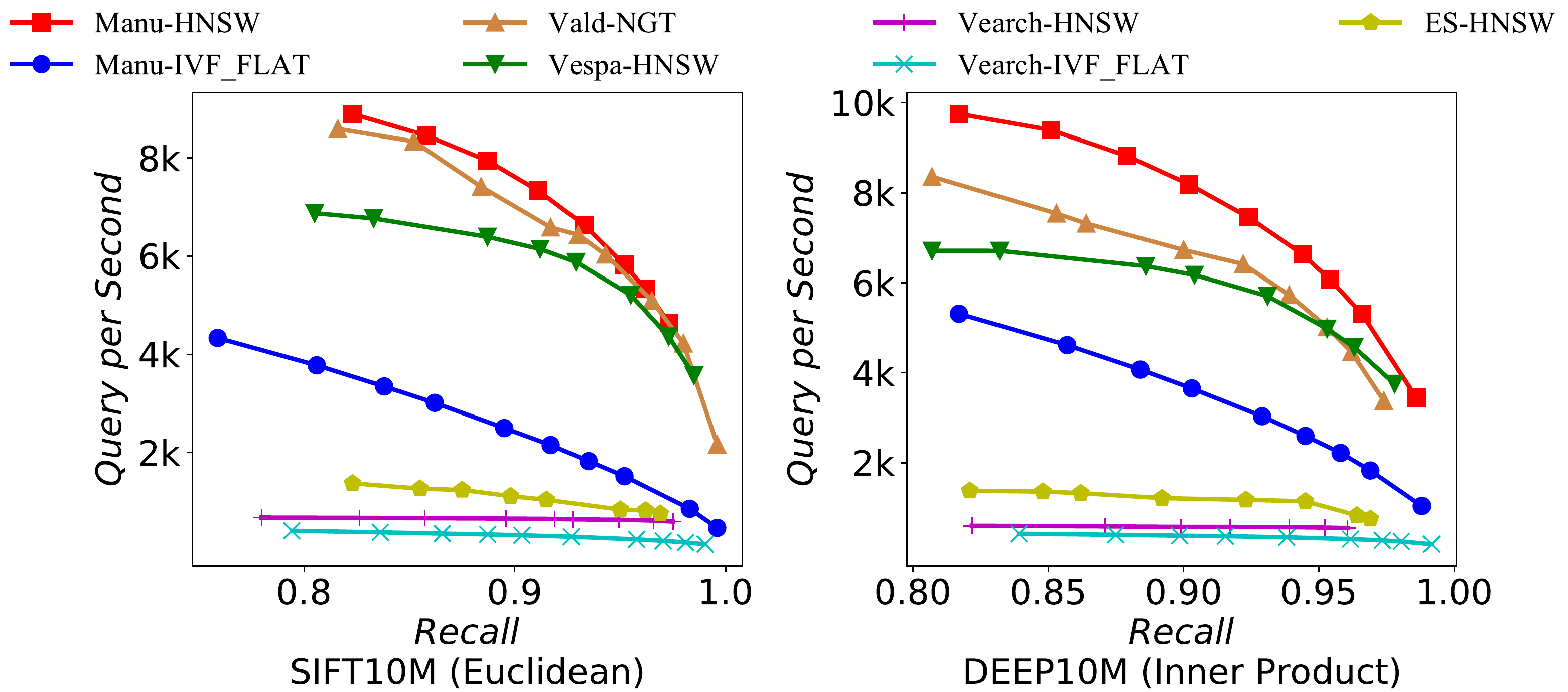}
	\vspace{-4mm}
	\caption{Recall vs. throughout comparison.}
	\vspace{-2mm}
	\label{exp:exp1-1}
	\vspace{-2mm}
\end{figure}

\sstitle{E-commerce recommendation} Company A is a leading online shopping platform in China that mainly sells clothing and makeups. They use \name~for recommendation, and products are recommended to a user according to their similarity scores with the user embedding vector. They have three main requirements for vector database: (1) \textit{high throughput} as they need to handle the requests of many concurrent costumers; (2) \textit{high quality} search results for good recommendation effect; (3) \textit{good elasticity} for low costs as their search requests have large fluctuations over time (peaks in evening but very low in midnight, very high at promotion events).




\begin{figure}[!t]	
	\centering 
	\includegraphics[width=0.7\linewidth]{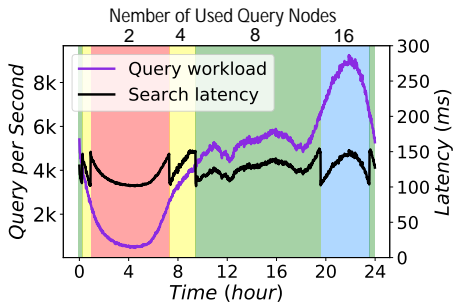}
	\vspace{-2mm}
	\caption{Search workload, query latency, and number of query nodes used by \name~over time. Different colors indicate different number of query nodes are used.}
	\vspace{-3mm}
	\label{exp:exp1-2}
	\vspace{-3mm}
\end{figure}


In Figure~\ref{exp:exp1-1}, we compare the recall-throughput performance of \name~with Elasticsearch (ES for short)~\cite{ref:es},  Vearch~\cite{li2018design}, Vald~\cite{vald}, and Vespa~\cite{vespa}, four popular open-source vector search systems, when using a single node. Note that the ES we use is the latest 8.0 version with tailored support for vector search instead of ES Plugin. We use Euclidean distance for SIFT and inner product for DEEP to test different similarity functions. Datasets with 10 million (10M) vectors are used as ES takes too much time to build index for larger dataset. As Vald only supports the NGT index~\cite{ngt} and Vespa only supports the HNSW index~\cite{malkov2018efficient} (both are efficient proximity graphs), we have only a single curve for them in each plot. The results show that \name~consistently outperforms the baselines across different datasets and similarity functions. ES and Vearch achieve significantly lower query processing throughput than \name~at the same recall. This is because that ES is a disk-based solution and Vearch's three-layer aggregation procedure (searcher-broker-blender) for search results introduces high overhead. The performances of Vald and Vespa are much better than ES and Vearch but still inferior compared with \name{}. We conjecture this is because \name{} has better implementations with optimizations for CPU cache and SIMD.

To test the elasticity of \name, we use the search traffic of an e-commerce platform over one day period~\cite{taobao}, which is plotted as the purple curve in Figure~\ref{exp:exp1-2}. The results show that search workload fluctuates violently over time, and the peak is much higher than the valley. We use SIFT100M as the dataset and Euclidean distance as the similarity function. \name~is configured to reduce query nodes by 0.5x when search latency is shorter than 100ms and add query nodes to 2x when search latency is over 150ms. The colors in Figure~\ref{exp:exp1-2} indicate the number of query nodes used by \name, which shows that \name~has good elasticity to adapt to query workload. The black line reports the search latency and shows that \name~can keep search latency within the target range via scaling.

\sstitle{Video deduplication}
Company B is a video sharing website in Europe, on which users can upload videos and share with others. They find that there are many duplicate videos that result in high management costs and thus conduct deduplication before archiving the videos. They model a video as a set of its critical frames and encode each frame into a vector. They use vector search to find videos in the corpus that are most similar to a new video and conduct further checking on the shortlisted videos to determine if the new video is a duplicate. They also use vector search to find videos similar to those viewed by users for recommendation. They require vector DBMS to have \textit{good scalability} with respect to both \textit{data volume} and \textit{computing resource} as their corpus grows quickly.

\begin{figure}[!t]	
	\centering 
	\includegraphics[width=1\linewidth]{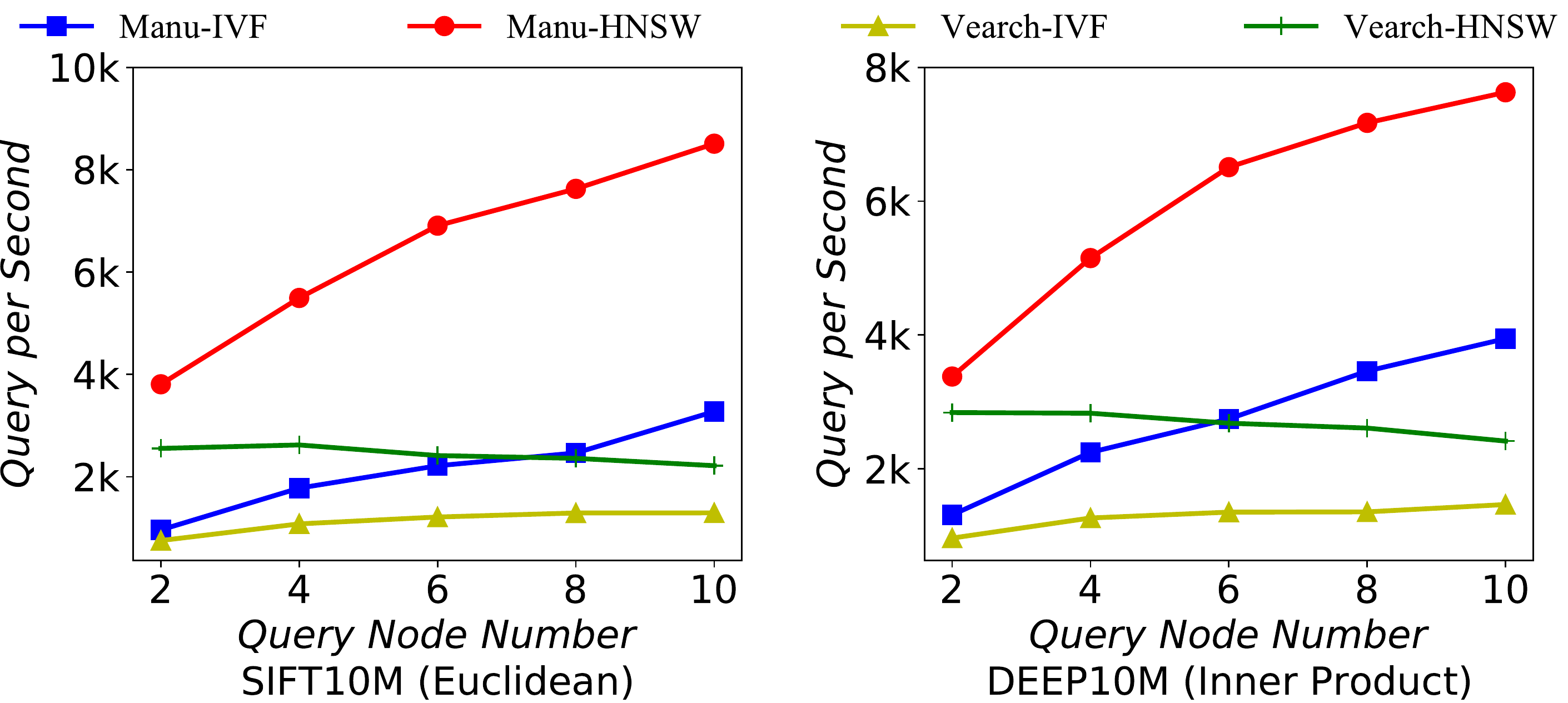}
	\vspace{-3mm}
	\caption{Scalability of \name~ w.r.t. query nodes.}
	\vspace{-3mm}
	\label{exp:exp3}
\end{figure}


\begin{figure}[!t]	
	\centering 
	\includegraphics[width=1\linewidth]{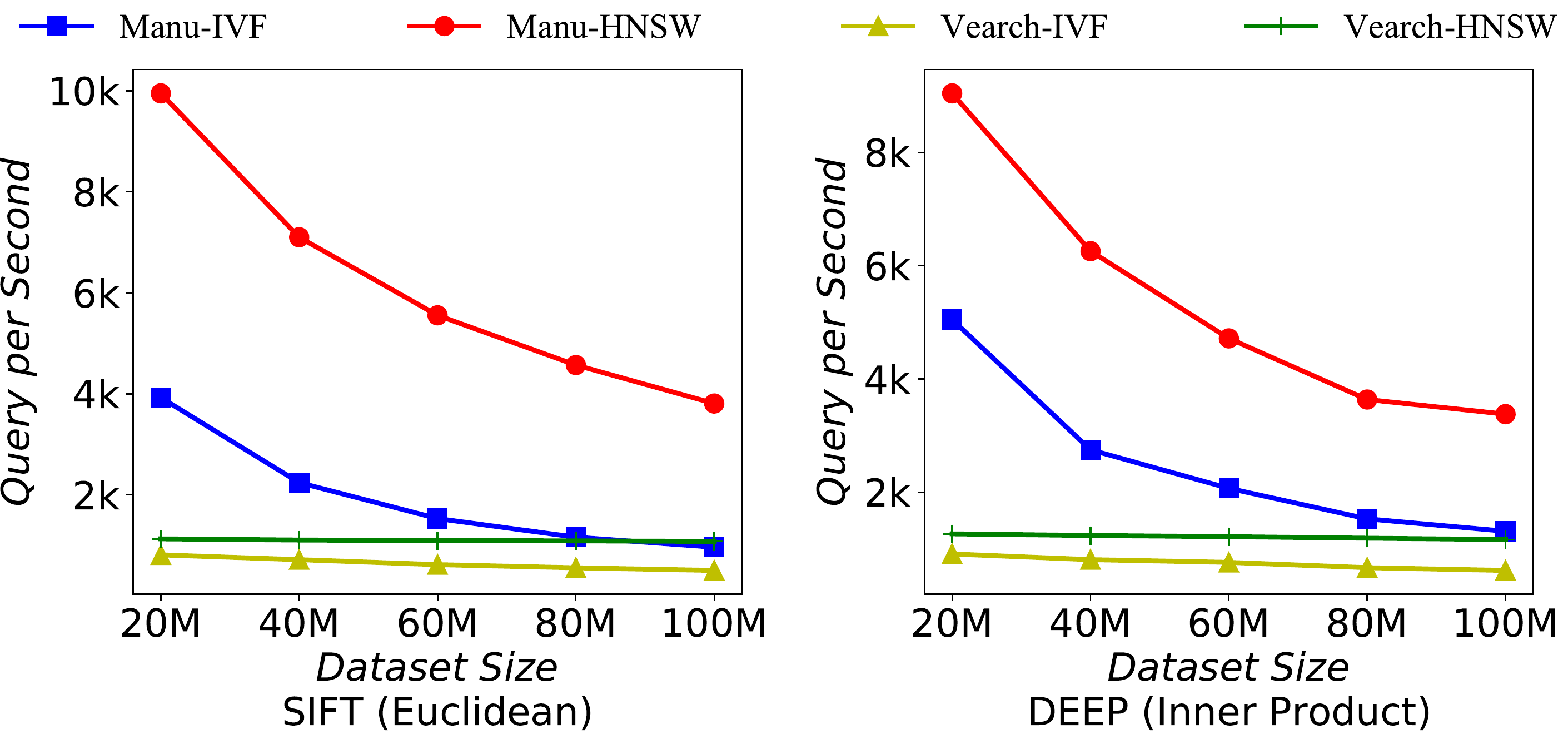}
	\vspace{-3mm}
	\caption{Scalability of \name~ w.r.t. data volume.}
	\vspace{-3mm}
	\label{exp:exp4}
\end{figure}


In Figure~\ref{exp:exp3} and Figure~\ref{exp:exp4}, we test the scalability of \name~when changing the number of query nodes and the size of dataset, respectively. The results show that query processing throughput scales almost linearly with the number of query nodes and the reciprocal of dataset size. The observation is consistent for different datasets, indexes and similarity functions. This is because \name~uses segments to distribute search tasks among query nodes. With segment size fixed, each query node handles more segments when data volume increases, and fewer segments when the number of query nodes increases. Note that better scalability w.r.t. data volume can be achieved by configuring \name~to use larger segments when dataset size increases. This is because similarity search indexes usually have sub-linear complexity w.r.t. dataset size.

\begin{figure}[!t]	
	\centering 
	\begin{minipage}[b]{0.23\textwidth}
		\includegraphics[width=\textwidth]{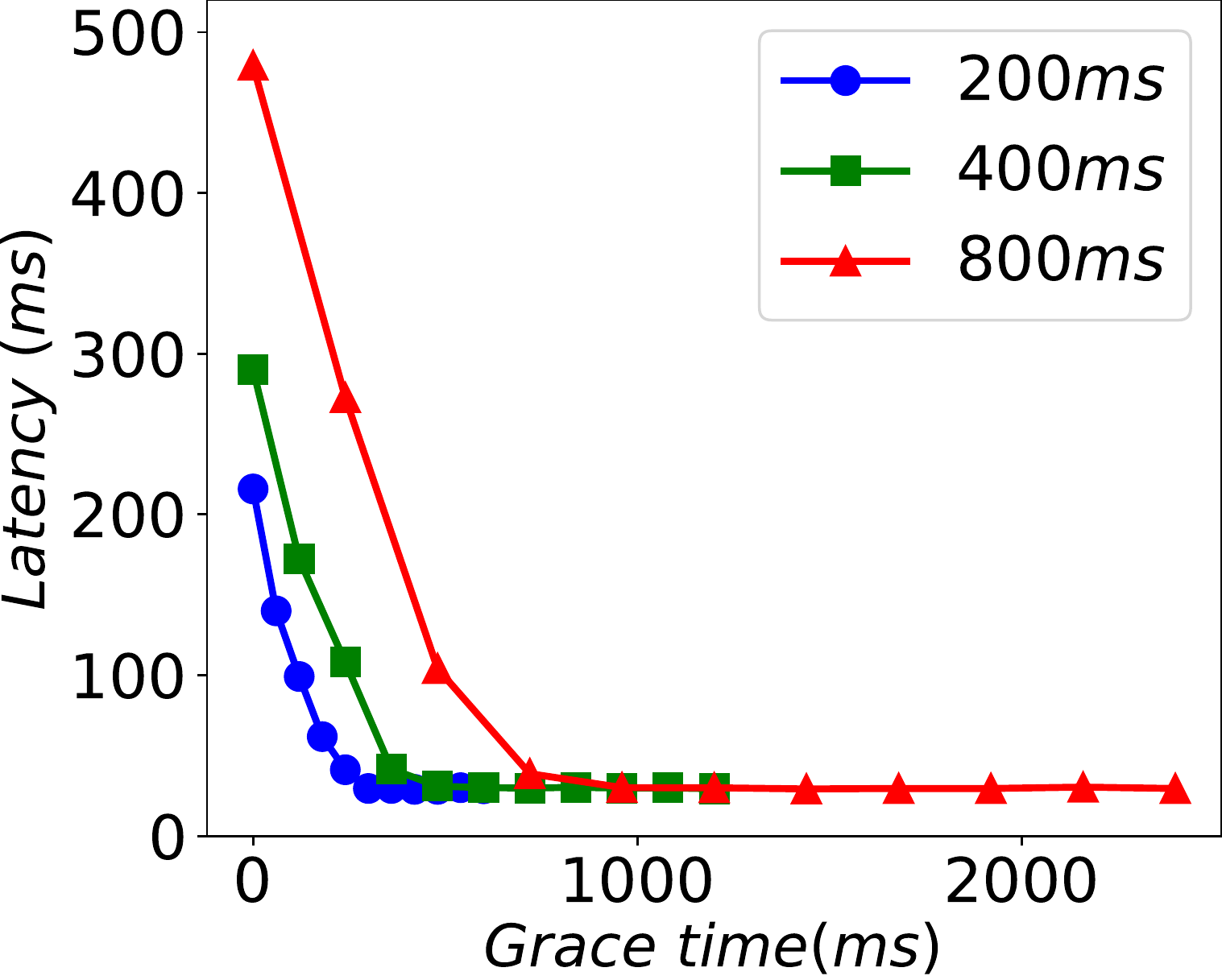}
		\subcaption{SIFT10M (Euclidean)}
	\end{minipage}
	\begin{minipage}[b]{0.23\textwidth}
		\includegraphics[width=\textwidth]{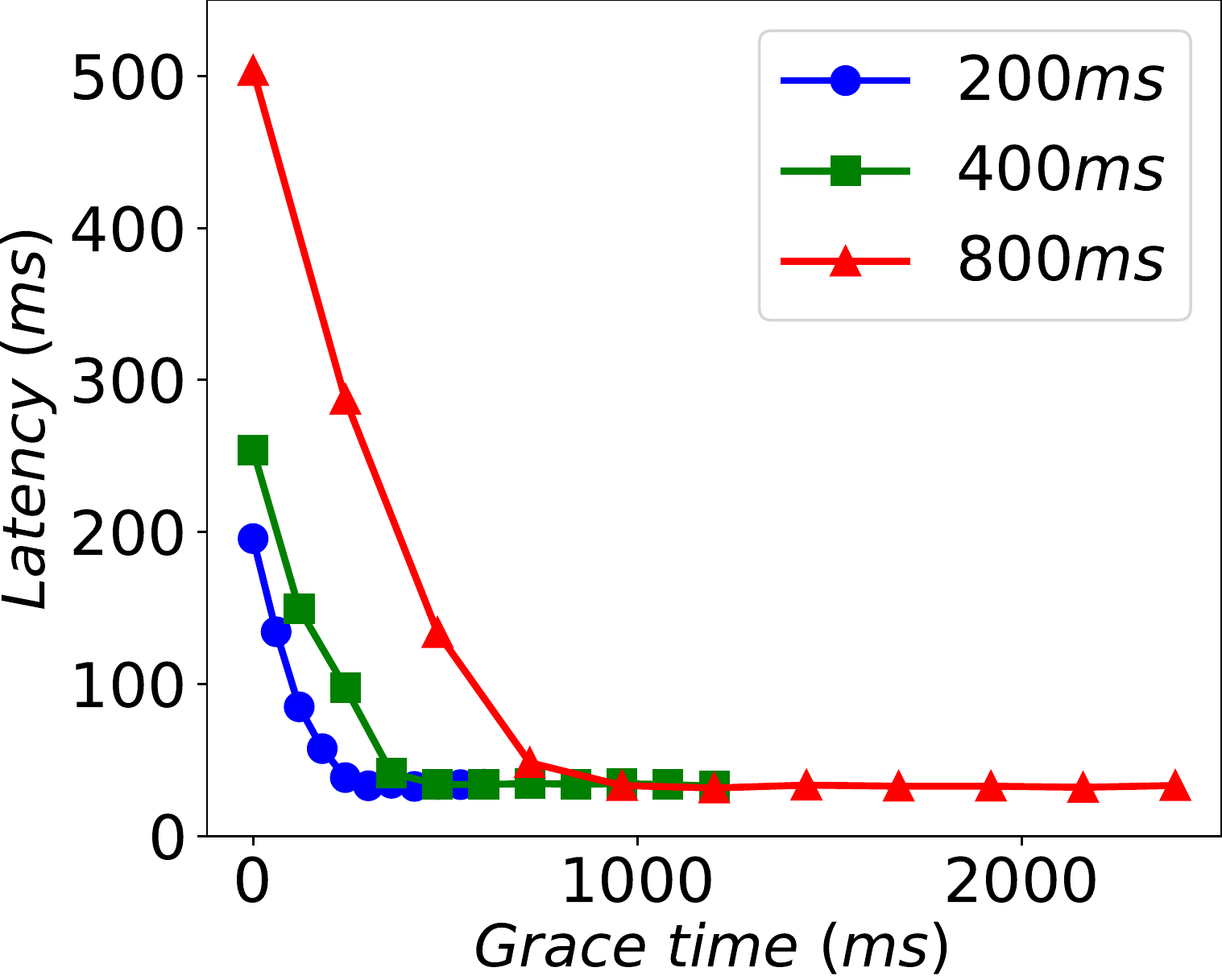}
		\subcaption{DEEP10M (Inner Product)}
	\end{minipage}
    \vspace{-2mm}
	\caption{The relation between search latency and grace time, legends stand for the time tick interval.} 
	\vspace{-2mm}
	\label{exp:exp5}
\end{figure}

\sstitle{Virus scan}
Company C is a world leading software security service provider and one of its main service is scanning viruses for smart phones. They have a virus base that continuously collects new viruses and develop specialized algorithms to map virus and user APK to vector embedding. To conduct a virus scan, they find viruses in their base that have embedding similar to the query APK and then compare the search results with the APK in more detail. They have two requirements for vector DBMS: (1) \textit{short delay} for streaming update as new viruses (vectors) are continuously added to their virus base and vector search needs to observe the latest viruses with a short delay; (2) \textit{fast index building} as they frequently adjust their embedding algorithm to fix problems, which leads to update of the entire dataset and requires to rebuild index.

\begin{figure}[!t]	
	\centering 
	\begin{minipage}[b]{0.23\textwidth}
		\includegraphics[width=\textwidth]{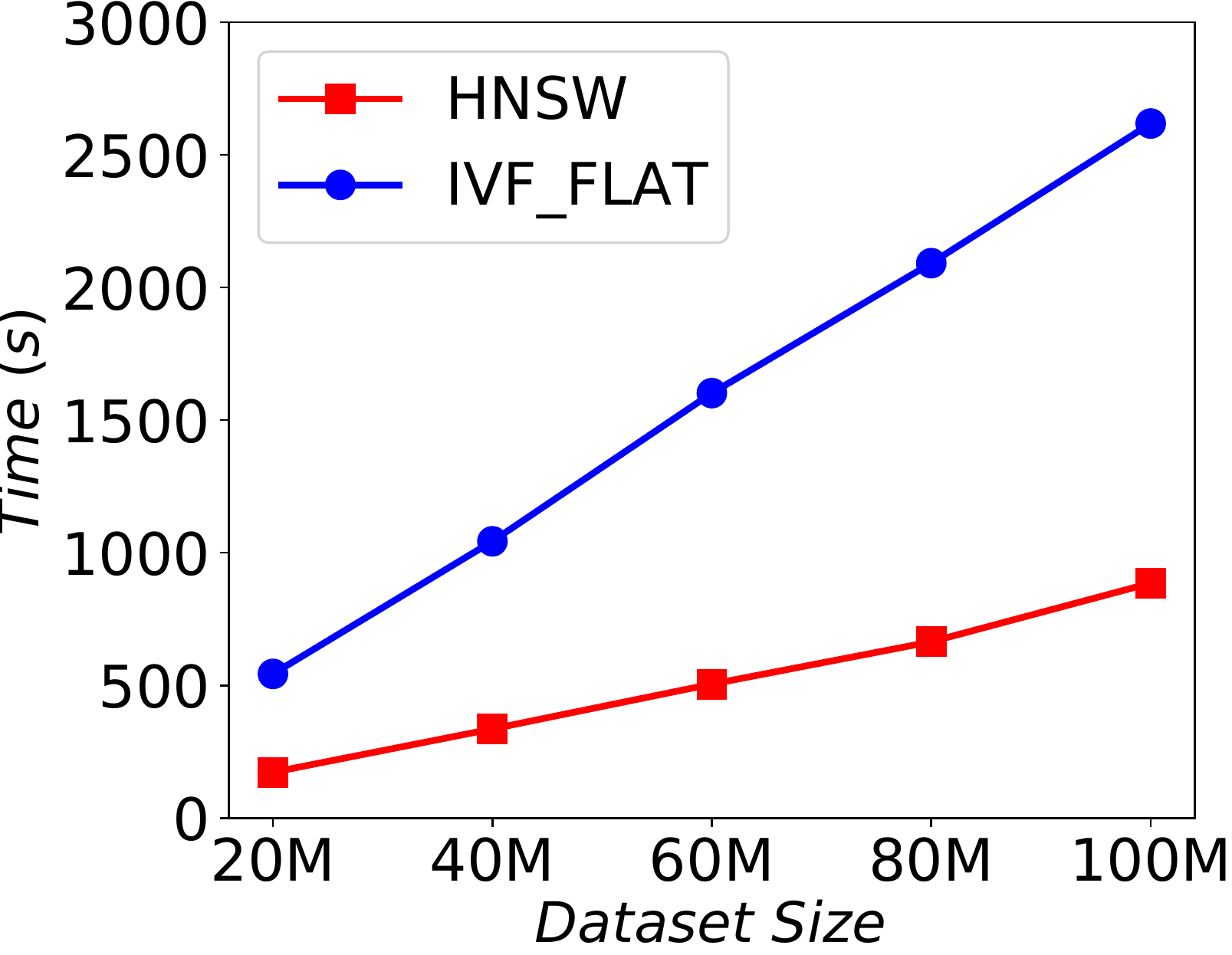}
		\subcaption{SIFT (Euclidean)}
	\end{minipage}
	\begin{minipage}[b]{0.23\textwidth}
		\includegraphics[width=\textwidth]{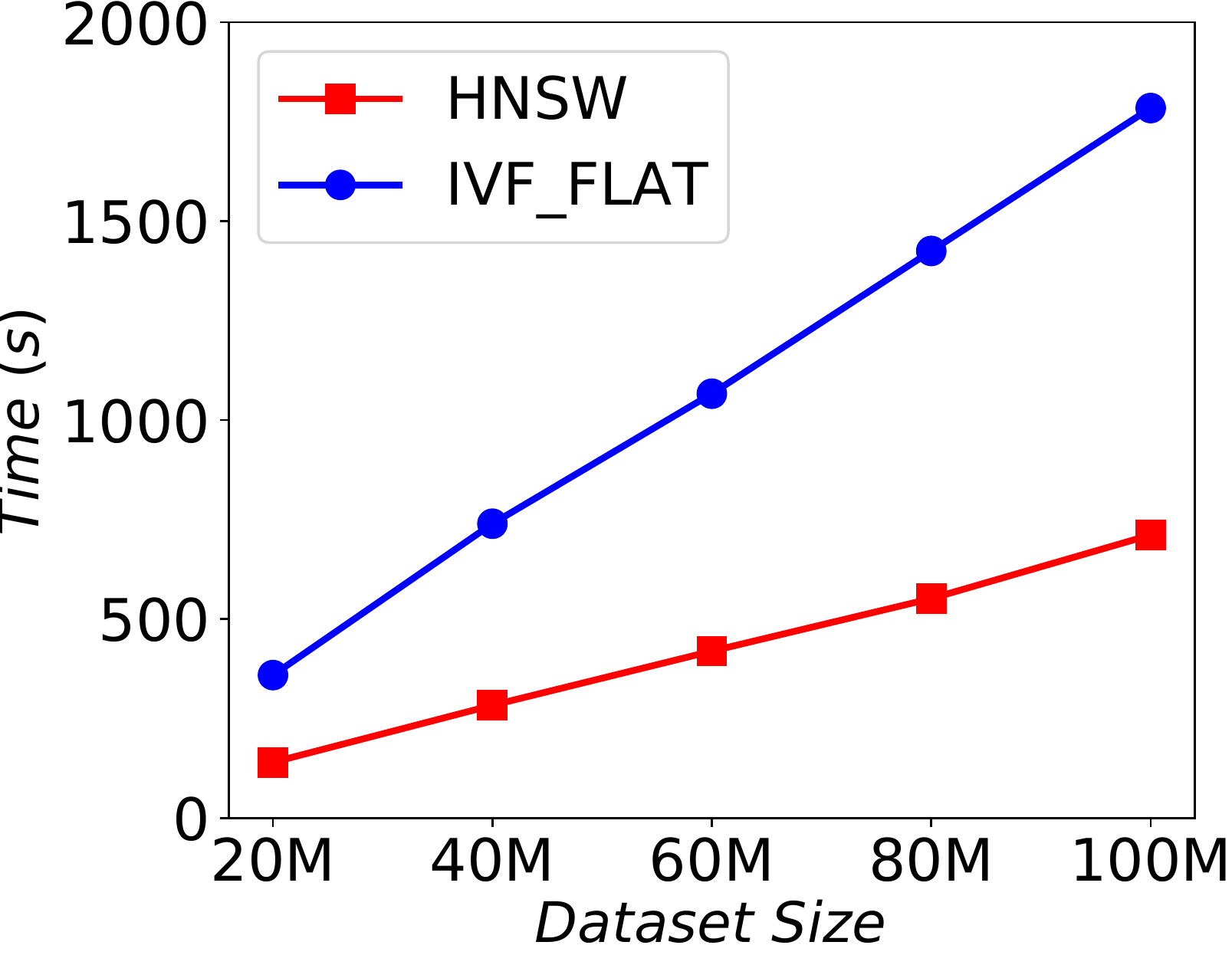}
		\subcaption{DEEP (Inner Product)}
	\end{minipage}
	\vspace{-2mm}	
	\caption{Index construction time of \name~vs. data volume.} 
	\vspace{-2mm}	
	\label{exp:exp6}
\end{figure}

In Figure~\ref{exp:exp5}, we show the average delay of search requests for \name. Recall that \textit{grace time} (i.e., $\tau$) means that a search request must observe updates that happen time $\tau$ before it, and is configurable by users. The legends correspond to different \textit{time tick interval}, with which the loggers write time tick to WAL. The results show that search latency decreases quickly with grace time, and shorter time tick interval results in shorter search latency. This is because with longer grace time, search requests can tolerate longer update delay and are less likely to wait for updates. When the time tick interval reduces, each segment can confirm that all updates have been received more quickly, thus the search requests wait for a shorter time. In Figure~\ref{exp:exp6}, we report the index building time of \name~when changing data volume. The results show that index building time scales linearly with data volume. This is because \name~builds index for each segment and larger data volume leads to more segments.

           

\section{Related Work}\label{sec:related}

\sstitle{Vector search algorithms}
Vector search algorithms have a long research history, and most works focus on efficient approximate search on large-scale datasets. Existing algorithms can be roughly classified into four categories, i.e., \textit{space partitioning tree} (SPT), \textit{locality sensitive hashing} (LSH), \textit{vector quantization} (VQ) and \textit{proximity graph} (PG). SPT algorithms divide the space into areas, and use tree structures to quickly narrow down search results to some areas~\cite{dasgupta2008random,silpa2008optimised,muja2014scalable,WangWJLZZH14, lu2020vhp}. LSH algorithms design hash functions such that similar vectors are hashed to the same bucket with high probability, and examples include~\cite{indyk1998approximate,shrivastava2014asymmetric,gionis1999similarity,gong2020idec,jafari2020mmlsh,li2020efficient,liu2019lsh,lu2020r2lsh,lv2017intelligent,zheng2020pm}. VQ algorithms compress vectors and accelerate similarity computation by quantizing the vectors using a set of vector codebooks, and well-known VQ algorithms include~\cite{jegou2010product,zhou2012scalar,iwasaki2018optimization, ge2013optimized,andre2016cache}. PG algorithms form a graph by connecting a vector with those most similar to it in the dataset, and conduct vector search by graph walk~\cite{malkov2018efficient, fu2019fast, SubramanyaDSKK19, ZhaoTL20}. Different algorithms have different trade-offs, e.g., LSH is cheap in indexing building but poor in result quality, VQ reduces memory and computation but also harms result quality, PG has high efficiency but requires large memory. \name{} supports a comprehensive set of search algorithms such that users can trade off between different factors.

\sstitle{Vector databases} Vector data management solutions have gone through two stages of development. Solutions in the first stage are libraries (e.g., Facebook Faiss~\cite{johnson2019billion}, Microsoft SPTAG~\cite{SPTAG}, HNSWlib~\cite{malkov2018efficient} and Annoy~\cite{ref:annoy}) and plugins (e.g., ES plugins~\cite{ref:es}, Postgres plugins~\cite{ref:postgresql}) for vector search.
They are insufficient for current applications as full-fledged management functionalities are required, e.g., distributed execution for scalability, online data update, and failure recovery. Two OLAP database systems, AnalyticDB-V~\cite{WeiWWLZ0C20} and PASE~\cite{YangLFW20} support vector data by adding a table column to store them but lacks optimizations tailored for vector data.

The second stage solutions are full-fledged vector databases such as Vearch~\cite{li2018design}, Vespa~\cite{vespa}, Weaviate~\cite{weaviate}, Vald~\cite{vald}, Qdrant~\cite{qdrant}, Pinecone~\cite{pinecone}, and our previous effort Milvus~\cite{wang2021milvus}.\footnote{Pinecone is offered as SaaS and closed source. Thus, we do not know its design details.} Vearch uses Faiss as the underlying search engine and adopts a three-layer aggregation procedure to conduct distributed search. Similarly, Vespa distributes data over nodes for scalability. A modified version of the HNSW algorithm is used to support online updates for vector data, and Vespa also allows attribute filtering during search and learning-based inference on search results (e.g., for re-ranking). Weaviate adopts a GraphQL interface and allows storing objects (e.g., texts, images), properties, and vectors. Users can directly import vectors or customize embedding models to map objects to vectors, and Weaviate can retrieve objects based on vector search results. 
Vald supports horizontal scalability by partitioning a vector dataset into segments and builds indexes without stopping search services.
Qdrant is a single-machine vector search engine with extensive support for attribute filtering. It allows filtering with various data types and query conditions (e.g., string matching, numerical ranges, geo-locations), and uses a tailored optimizer to determine the filtering strategy. Note that Vespa, Weaviate and Vald only support proximity graph index.

We can observe that these vector databases focus on different functionalities, e.g., learning-based inference, embedding generation, object retrieval, and attribute filtering. Thus, we treat evolvability as first class priority when design \name{} such that new functionalities can be easily introduced.   
\name{} also differs from these vector databases in important perspectives. First, the log backbone of \name{} provides time semantics and allows tunable consistency. Second, \name{} decomposes system functionalities with fine granularity and instantiates them as cloud services for performance and failure isolation, and thus is more suitable for cloud deployment. Third, \name{} comes with more comprehensive optimizations for usability and performance, e.g., support various indexes, hardware tailored implementations, and GUI tools.

\sstitle{Cloud native databases} Many OLAP databases are designed to run on the cloud recently and examples include Redshift~\cite{GuptaATKPSS15}, BigQuery~\cite{MelnikGLRSTV10}, Snowflake~\cite{DagevilleCZAABC16} and AnalyticDB~\cite{zhan2019analyticdb}. Redshift is a data warehouse system offered as a service on Amazon Web Service and adopts a \textit{shared-nothing} architecture. It scales by adding or removing EC2 instances, and data is redistributed in the granularity of columns. Snowflake uses a \textit{shared-data} architecture by delegating data storage to Amazon S3. Compute nodes are stateless and fetch read-only copies of data for tasks, and thus can be easily scaled. For efficiency, high-performance local disk is used to cache hot data.

Aurora~\cite{VerbitskiGSBGMK17} and PolarDB Serverless~\cite{Li19} are two cloud native OLTP databases. Aurora uses a \textit{shared-disk} architecture and proposes the ``log is database" principle by pushing transaction processing down to the storage engine. It observes that the bottleneck of cloud-based platforms has shifted from computation and storage IO to network IO. Thus, it only persists redo log for transaction processing and commits transaction by processing log according to LSN. PolarDB Serverless adopts a \textit{disaggregation} architecture, which uses high-speed RDMA network to decouple hardware resources (e.g., compute, memory and storage) as resource pools.

Our \name~follows the general design principles of cloud native databases to decouple the system functionalities at fine granularity for high elasticity, fast evolution and failure isolation. However, we also consider the unique design opportunities of vector databases to trade the simple data model and weak consistency requirement for performance, cost and flexibility. Specifically, complex transactions are not supported and the log backbone is utilized to support tunable consistency-performance trade-off. Moreover, vector search, index building and log archiving tasks are further decoupled as their workloads may have significant variations. 
\section{Conclusions and Future Directions}\label{sec:conclusion}


In this paper, we introduce the designs of \name~as a cloud native vector database. To ensure that \name~suits vector data applications, we set ambitious design goals, which include good evolvability, tunable consistency, high elasticity, good efficiency, and etc. To meet these design goals, \name~trades the simple data model of vectors and weak consistency requirement of applications for performance, costs and flexibility. Specifically, \name~conducts fine-grained decoupling of the system functionalities for component-wise scaling and evolution, and uses the log backbone to connect the system components while providing time semantics and simplifying inter-component interaction. We also introduce important features such as high-level API, GUI tool, hardware optimizations, and complex search. We think \name~is still far from perfect and some of our future directions include:


\squishlist

\item \stitle{Multi-way search} Many applications jointly search multiple types of contents, e.g., vector and primary key, vector and text. The log system of \name~allows to add search engines for other contents (e.g., primary key and text) as co-processors by subscribing to the log stream. We will explore how multiple search engines can interact efficiently and how to flexibly coordinate different search engines to meet application requirements.  
  
\item \stitle{Modularized algorithms} We think vector search algorithms can be distilled into independent components, e.g., \textit{compression} for memory reduction and efficient computation, \textit{indexing} for limiting computation to a small portion of vectors, and \textit{bucketing} for grouping similar vectors. Existing vector search algorithms only explore some combinations of techniques for different components. We will provide a unified framework for vector search such that users can flexibly combine different techniques according to their desired trade-off between cost and performance.

\item \stitle{Hierarchical storage aware index} Current vector search index assumes a single type of storage, e.g., GPU memory, main memory or disk. We will explore indexes that can jointly utilize all devices on the storage hierarchy. For example, most applications have some hot vectors (e.g., popular products in e-commerce) that are frequently accessed by search requests, which can be placed in fast storage. As a query accesses only a portion of the vectors and a node processes many concurrent queries, the storage swap latency may be hidden by pipelining.   

\item \stitle{Advanced hardware} NVM~\cite{ren2020hm} costs about one-third of DRAM for unit capacity but provides comparable read bandwidth and latency comparable, which makes it a good choice for replacing expensive DRAM when storing large datasets. RDMA~\cite{kalia2014using,cao2021polardb} significantly reduces the communication latency among nodes, and NVLink~\cite{pearson2018numa} directly connects GPUs with much larger bandwidth than PCIe. By exploiting these fast connections, we will explore indexes and search algorithms that jointly use multiple devices. We are also working with hardware vendors to apply FPGA and MLU for vector search and index building.

\item \stitle{Embedding generation toolbox} For better application level-integration, we plan to incorporate a application-oriented toolbox for generating embedding vectors. This toolbox would incorporate model fine-tuning in addition to providing a number of pre-trained models that can be used out-of-the-box, allowing for rapid prototyping.

\squishend

\begin{acks}
\name~is a multi-year project open sourced by Zilliz. The development of \name~involves many engineers in its community. In particular, we thank Bingyi Sun, Weida Zhu, Yudong Cai, Yihua Mo, Xi Ge, Yihao Dai, Jiquan Long, Cai Zhang, Congqi Xia, Xuan Yang, Binbin Lv, Xiaoyun Liu, Wenxing Zhu, Yufen Zong, Jie Zeng, Shaoyue Chen, Jing Li, Zizhao Chen, Jialian Ji, Min Tian, Yan Wang and all the other contributors in the community for their contributions. We also thank Filip Haltmayer for proofreading the paper and valuable suggestions to improve paper quality.
\end{acks}

\bibliographystyle{ACM-Reference-Format}
\bibliography{ref.bib}

\end{document}